\documentclass[sn-mathphys,Numbered,pdflatex,iicol]{sn-jnl}% Math and Physical Sciences Reference Style
\usepackage{graphicx}%
\usepackage{multirow}%
\usepackage{amsmath,amssymb,amsfonts}%
\usepackage{amsthm}%
\usepackage{mathrsfs}%
\usepackage[title]{appendix}%
\usepackage{xcolor}%
\usepackage{textcomp}%
\usepackage{manyfoot}%
\usepackage{booktabs}%
\usepackage{algorithm}%
\usepackage{algorithmicx}%
\usepackage{algpseudocode}%
\usepackage{listings}%
\usepackage{bm}%
\usepackage{url}%
\usepackage[caption=false]{subfig}%
\usepackage{comment}%
\usepackage{dsfont}%
\usepackage{flushend}%
\usepackage{multicol}

\hypersetup{
    colorlinks,
    linkcolor={black!50!black},
    citecolor={black!50!black},
    urlcolor={red!50!red}
}

\begin{document}

\title[Data-Driven Room Acoustic Modeling Via Differentiable Feedback Delay Networks With Learnable Delay Lines]{Data-Driven Room Acoustic Modeling Via Differentiable Feedback Delay Networks With Learnable Delay Lines}


\author*[1]{\fnm{Alessandro Ilic} \sur{Mezza}}\email{alessandroilic.mezza@polimi.it}

\author[1]{\fnm{Riccardo} \sur{Giampiccolo}}\email{riccardo.giampiccolo@polimi.it}

\author[2]{\fnm{Enzo} \sur{De Sena}}\email{e.desena@surrey.ac.uk}

\author[1]{\fnm{Alberto} \sur{Bernardini}}\email{alberto.bernardini@polimi.it}

\affil*[1]{\orgdiv{Dipartimento di Elettronica, Informazione e Bioingegneria}, \orgname{Politecnico di Milano}, \orgaddress{\street{Piazza Leonardo da Vinci, 32}, \city{Milan}, \postcode{20133}, \country{Italy}}}

\affil[2]{\orgdiv{Institute of Sound Recording}, \orgname{University of Surrey}, \orgaddress{\street{Stag Hill, University Campus}, \city{Guildford}, \postcode{GU27XH}, \country{UK}}}


\abstract{%
Over the past few decades, extensive research has been devoted to the design of artificial reverberation algorithms aimed at emulating the room acoustics of physical environments. Despite significant advancements, automatic parameter tuning of delay-network models remains an open challenge. We introduce a novel method for finding the parameters of a Feedback Delay Network (FDN) such that its output renders target attributes of a measured room impulse response. The proposed approach involves the implementation of a differentiable FDN with trainable delay lines, which, for the first time, allows us to simultaneously learn each and every delay-network parameter via backpropagation. The iterative optimization process seeks to minimize a perceptually-motivated time-domain loss function incorporating differentiable terms accounting for energy decay and echo density. Through experimental validation, we show that the proposed method yields time-invariant frequency-independent FDNs capable of closely matching the desired acoustical characteristics, and outperforms existing methods based on genetic algorithms and analytical FDN design.
}%



\onecolumn

\centerline{\huge Data-Driven Room Acoustic Modeling Via Differentiable}
\centerline{\huge Feedback Delay Networks With Learnable Delay Lines}
\vspace{1.5em}
\centerline{\large Alessandro Ilic Mezza$^1$, Riccardo Giampiccolo$^1$, Enzo De Sena$^2$, and Alberto Bernardini$^1$}
\vspace{0.5em}
\centerline{$^1$ Dipartimento di Elettronica, Informazione e Bioingegneria, Politecnico di Milano,}
\centerline{Piazza Leonardo da Vinci, 32, Milan, 20133, Italy}
\centerline{$^2$ Institute of Sound Recording, University of Surrey,}
\centerline{Stag Hill, University Campus, Guildford, GU27XH, UK}
\vspace{4.5em}

    \centerline{\Large\bf \color{red} Peer-Reviewed Version}
    \vspace{0.5em}
    
    {\color{red} \noindent A peer-reviewed version of the manuscript has been published in \textit{EURASIP Journal of Audio, Speech, and Music Processing} and is available at \href{http://dx.doi.org/10.1186/s13636-024-00371-5}{10.1186/s13636-024-00371-5}. \bf Please, cite this work as:\\ 
    
    A. I. Mezza, R. Giampiccolo, E. De Sena, and A. Bernardini, ``Data-Driven Room Acoustic Modeling Via Differentiable Feedback Delay Networks With Learnable Delay Lines," \textit{EURASIP Journal of Audio, Speech, and Music Processing}, vol. 2024, no. 1, pp. 1-20 (51), 2024, doi: 10.1186/s13636-024-00371-5.\\
    
    \noindent\texttt{@article\{mezza2024dfdn,\\
    title = \{Data-Driven Room Acoustic Modeling Via Differentiable Feedback Delay Networks With Learnable Delay Lines\},\\
    author = \{Mezza, Alessandro Ilic  and Giampiccolo, Riccardo and De Sena, Enzo and Bernardini, Alberto\},\\
    journal = \{EURASIP Journal of Audio, Speech, and Music Processing\},\\
    volume = \{2024\},\\
    number = \{1\}, \\
    pages = \{1-20\},\\
    year = \{2024\},\\
    issn = \{1687-4722\},\\
    doi = \{https://doi.org/10.1186/s13636-024-00371-5\}\\
    \}}
    }%

\hypersetup{
    colorlinks,
    linkcolor={black!50!black},
    citecolor={black!50!black},
    urlcolor={black!50!black}
}

\maketitle

\twocolumn
\section{Introduction}
\label{sec:introduction}
Room acoustic synthesis involves simulating the acoustic response of an environment, a task that finds application in a variety of fields, e.g., in music production, to artistically enhance sound recordings; in architectural acoustics, to improve the acoustics of performance spaces; or in VR/AR/computer games, to enhance listeners’ sense of realism~\cite{apostolopoulos2012road}, immersion~\cite{potter2022relative}, and externalization~\cite{geronazzo2020minimal}.

Room acoustic models can be broadly classified in physical models, convolution models, and delay-network models~\cite{fifty_years}. 
Physical ones can be further divided in wave-based models, which provide high physical accuracy but at the cost of significant computational complexity, and geometrical-based ones, which make the simplifying approximation that sound travels like rays. 
Convolution models involve a set of stored room impulse responses (RIRs) and are therefore capable of replicating the true response of a real room~\cite{fifty_years}.
Convolution is, however, an operation that despite recent advances~\cite{wefers2015partitioned} still carries a computational load that makes it ill-suited in certain real-time applications. 

Delay-network models consist of recursively connected networks of delay lines and have a significantly lower computational cost than convolution. 
Rather than modeling the physical response of a specific room, delay-network models only aim to replicate certain perceptual aspects of room acoustics. 
These models have a long history, which can be traced back to the Schroeder reverberator~\cite{schroeder_natural}.
Since then, a number of designs have been proposed, including feedback delay networks (FDNs)~\cite{jot_fdn,unilossless, fdn_art}, scattering delay networks (SDNs)~\cite{efficient_sdn}, and waveguide networks (WGNs)~\cite{wgw}.

The parameters of delay-network models are typically designed to obtain certain desired acoustical characteristics, e.g., a target reverberation time. 
An alternative design paradigm is to fit the parameters such that the output is as close as possible to that of a measured RIR, hence combining the accuracy of convolution models with the low computational complexity of delay-network models. 
Several methods following this alternative design paradigm have been recently proposed for the case of FDNs, for instance using gradient-free methods~\cite{bona2022avanzini, chemistruck2012generating, shen2020data, coggin2016automatic, ibnyahya2022method} and 
gradient-based machine learning techniques~\cite{lee2022differentiable, dalsanto2023colorless}. 
Existing approaches, however, involve a certain degree of human intervention and require heuristic-driven ad-hoc choices for several model parameters.

This paper proposes a new method for automatic FDN parameter tuning. 
The present work is rooted in a recent framework for the parameter estimation of lumped-element models \cite{mezza2023lumped} based on automatic differentiation~\cite{baydin2018automatic}, and its novelty is twofold.
First, the cost function combines two objective measures of perceptual features, i.e., the Energy Decay Curve (EDC) and a differentiable version of the normalized Echo Density Profile (EDP)~\cite{abel2006simple}.
Second, the delay line lengths are optimized via backpropagation along with every other FDN parameter, thus allowing exploiting the flexibility of delay-network models to the fullest. 
{\color{black}
We thus introduce a simple, robust, and fully automatic method for matching acoustic measurements.}
The learned parameters can be then seamlessly plugged into off-the-shelf FDN software without further processing or mapping.

The paper is organized as follows. Section~\ref{sec:fdn} introduces the background information on FDNs. Section~\ref{sec:related_work} discusses the prior art on automatic FDN parameter tuning. Section~\ref{sec:proposed_method} describes the proposed method, and Section~\ref{sec:evaluation} presents its evaluation. Finally, Section~\ref{ref:conclusions} concludes the manuscript.

\begin{figure}[t]
    \centering
    \includegraphics[width=\linewidth]{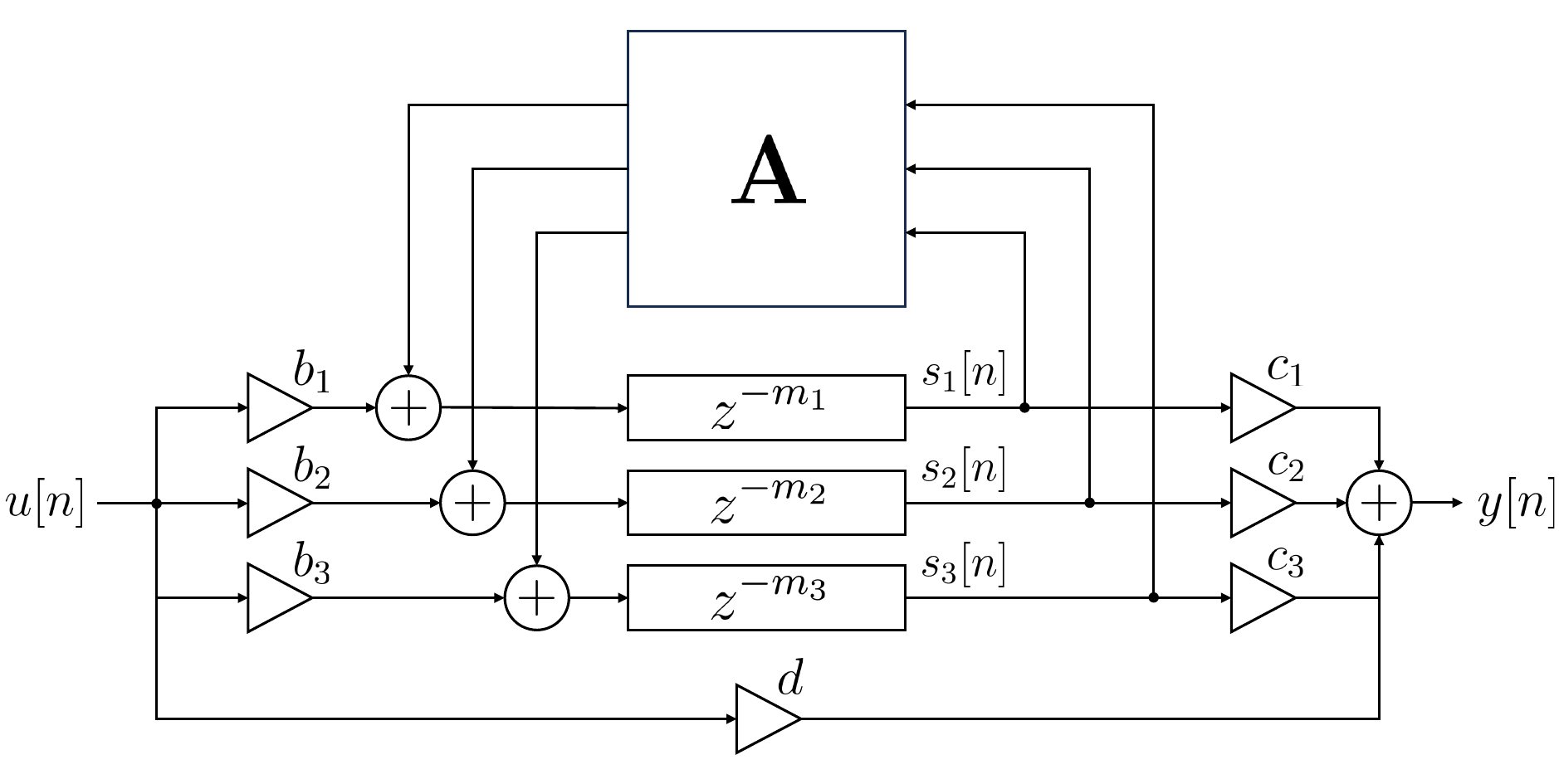}
    \caption{Block diagram of a SISO FDN with $N=3$.}
    \label{fig:siso_fdn}
\end{figure}

\section{Feedback Delay Networks}
\label{sec:fdn}
The block diagram of a single-input-single-output (SISO) FDN is shown in Figure~\ref{fig:siso_fdn}.
This system is characterized by~\cite{rocchesso1997fdn}
\begin{equation}
    \begin{array}{rl}
     y[n] &= \mathbf{c}^T\mathbf{s}[n]  + d u[n] \\
     \mathbf{s}[n + \bm{m}] &= \mathbf{A}\;\mathbf{s}[n] + \mathbf{b}u[n],
\end{array}
\label{eq:time_domain_fdn}
\end{equation}
where $u[n]$ is the input signal, $y[n]$ is the output signal, $\mathbf{b}\in\mathbb{R}^N$ is a vector of input gains, $\mathbf{c}\in\mathbb{R}^N$ is a vector of output gains, $(\cdot)^T$ denotes the transpose operation, $\mathbf{A}\in\mathbb{R}^{N\times N}$ is the feedback matrix, $d\in\mathbb{R}$ is the scalar gain associated to the direct path, and $\bm{m}=[m_1, ..., m_N]$ is a vector containing the length of the $N$ delay lines expressed in samples. The vector $\mathbf{s}[n]\in\mathbb{R}^N$ denotes the output of the delay lines at time index $n$, and we use the following notation $\mathbf{s}[n+\bm{m}]=\left[s_1[n + m_1], ..., s_N[n + m_N]\right]^T$ to indicate $N$ parallel delay operations of $m_1, ..., m_N$ samples, respectively, applied to $\mathbf{s}[n]$.

If $\bm{m}=[1,...,1]$, then \eqref{eq:time_domain_fdn} corresponds to the measurement and state equations of a state-space model.
In other words, an FDN corresponds to a generalized version of a state-space model with non-unit delays~\cite{rocchesso1997fdn}. 

The standard approach to designing the FDN parameters involves choosing the feedback matrix, delays, and input/output weights so as to obtain certain desired acoustic characteristics---usually a sufficient echo density and a pre-set reverberation time. 
The most important parameters are the ones associated to the recursive loop, i.e., $\bm{m}$ and $\mathbf{A}$, since they determine the energy decay behavior of the model, as well as its stability. 
The delays, $\bm{m}$, are typically chosen as co-prime of each other, so as to reduce the number of overlapping echoes and increase the echo density~\cite{schlecht2016feedback}. 
The design of the feedback matrix, $\mathbf{A}$, starts from a lossless prototype, usually an orthogonal matrix such as Hadamard or Householder matrix, which have been shown to ensure (critical) stability regardless of the delays, a property defined by Schlecht and Habets as \emph{unilosslessness}~\cite{unilossless}.
Losses are then incorporated by multiplying the unilossless matrix by a diagonal matrix of scalars designed to achieve a pre-set reverberation time, $T_{60}$. 

While it is possible to design feedback matrices as time varying and/or frequency dependent~\cite{schlecht2015time,jot_fdn}, this paper focuses on the time-invariant and frequency-independent case. 
With this assumption, the stability of the system can be easily enforced throughout the training, thanks to the model reparameterization strategies discussed later in Section~\ref{sec:proposed_method}. 
Moreover, 
{\color{black} time-invariant frequency-independent FDNs benefit from having a low computational complexity.}
From visual inspection of Figure~\ref{fig:siso_fdn}, indeed, such an FDN only requires $2N+1$ multiplications, $2N$ additions and one vector-matrix multiplication per sample.
The vector-matrix multiplication requires $N^2$ scalar multiplications and $N(N-1)$ additions for the case of a generic feedback matrix (while it becomes $O(N)$ for a Householder matrix). 
Assuming equal cost of additions and multiplications, the overall computational complexity of an FDN amounts to $f_\mathrm{s}\:
(2N^2+3N+1)$ floating-point operations per second (FLOPS), where $f_\mathrm{s}$ is the sampling rate. 
For $N=6$ and $f_\mathrm{s}=44.1$~kHz, that corresponds to a computational complexity of $4$~MFLOPS.
For comparison, modeling a $0.5$ s long RIR using an FIR filter (i.e., naive convolution) at the same sampling rate would carry a computation complexity of $1945$~MFLOPS.
In real-time applications, one would normally use faster methods such as partitioned convolution~\cite{wefers2015partitioned} or overlap-add (FFT-based) convolution~\cite{oppenheim1989discrete}.
Under the same conditions and assuming a frame refresh rate of $50$~Hz, overlap-add convolution carries a complexity of $207$~MFLOPS ~\cite{efficient_sdn}, which is still nearly two orders of magnitude larger than an FDN. 

\section{Related Work}
\label{sec:related_work}
As mentioned earlier, the automatic tuning of FDN parameters has been previously investigated by means of gradient-free methods, such as Bayesian optimization~\cite{bona2022avanzini} and genetic algorithms~\cite{chemistruck2012generating, coggin2016automatic, shen2020data, ibnyahya2022method}, as well as 
gradient-based machine learning techniques~\cite{lee2022differentiable, dalsanto2023colorless}. 

Some works are concerned with the automatic tuning of off-the-shelf reverberation plug-ins.
In~\cite{heise2009automatic}, Heise and colleagues investigate four gradient-free optimization strategies: simulated evolution~\cite{fogel1999intelligence}, the Nelder-Mead simplex method~\cite{nelder1965simplex}, Nelder-Mead with brute-force parallelization, and particle swarm optimization~\cite{kennedy1995particle}.
More recently, \cite{bona2022avanzini}~applies Bayesian optimization using a Gaussian process as a
prior to iteratively acquire the control parameters of an external FDN plug-in that minimize the mean absolute error between the multiresolution mel-spectrogram of the target RIR convolved with a 3~s logarithmic sine sweep and that of the artificial reverberator output. The FDN control parameters include the delay line length, reverberation time, fade-in time,
high/low cutoff, high/low Q, high/low gain, and dry-wet ratio.

Conversely, other studies assume to have white-box access to the delay-network structure and apply genetic algorithms (GA) to optimize a subset of the FDN parameters. 
In~\cite{chemistruck2012generating}, a GA is used to find both the $N^2$ coefficients of the feedback matrix $\mathbf{A}$ and $N$ cutoff frequencies of lowpass filters, one for each delay line. 
The authors of~\cite{shen2020data} aim at finding a mapping between room and FDN parameters for VR/AR applications. To this end, they synthesize the binaural RIRs of a set of virtual shoebox rooms, apply a GA to tune the FDN's delay lines and scalar feedback gain, and use the resulting training pairs to fit a support vector machine (SVM) regressor.  
In \cite{coggin2016automatic}, Coggin and Pirkle apply a GA for the estimation of  $\bm{m}$, $\mathbf{b}$, and $\mathbf{c}$. For every individual in a generation, attenuation and output filters are designed using the Yule-Walker method. The authors investigate several fitness functions before favoring the Chebyshev distance between the power envelopes of the target and predicted IR. The optimization is run for late reverberation only: the first $85$~ms of the RIR are cut and convolved with the input signal, before being fed to the FDN to model late reverberation.
Following~\cite{coggin2016automatic}, Ibnyahya and Reiss recently introduced a multi-stage method~\cite{ibnyahya2022method} combining more advanced analytical filter design methods and GAs to estimate the FDN parameters that would best approximate a target RIR in terms of an MFCC-based fitness function similar to the cost function used in~\cite{heise2009automatic}.

Due to the well-known limitations of genetic algorithms, such as the high risk of finding sub-optimal solutions, overall slow convergence rate, and the challenges of striking a good exploration-exploitation balance~\cite{vcrepinvsek2013exploration},
gradient-based techniques 
%inspired by differentiable machine learning 
have been recently proposed.

Inspired by groundbreaking research on differentiable digital signal processing~\cite{engel2020ddsp}, Lee et al.~\cite{lee2022differentiable} let the gradients of a  multiresolution spectral loss flow through a differentiable artificial reverberator so that they may reach a trainable neural network tasked with yielding the reverberator parameters.
This way, the authors train a convolutional-recurrent neural network tasked with inferring the input, output, and absorption filters of a FDN from a reference reverberation (RIR or speech).
It is worth mentioning, however, that it is not the delay-network parameters those that are optimized via stochastic gradient descent, but rather it is the weights of the neural network serving as black-box parameter estimator.
As such, the differentiable FDN is effectively used as a processing block in computing the loss of an end-to-end neural network instead of being the target of the optimization process. 

In a different vein, several recent works aim at learning lumped parameters via gradient-based optimization directly within the digital structure of the model and forgo parameter-yielding neural networks altogether.
In this respect, automatic differentiation has been recently proposed to find  $\mathbf{A}, \mathbf{b}, and \mathbf{c}$ of an FDN (without parameterizing them as a neural network) so as to minimize spectral coloration and obtain a flat frequency response~\cite{dalsanto2023colorless}.
Similar yet distinct, other works adopt a white-box system identification approach and use backpropagation to find the parameters of predetermined mathematical models so as to match measured data as closely as possible~\cite{esqueda2021whitebox, shintani2022mosfet, mezza2023lumped}.

In this work, we adopt the latter approach and use the method detailed in the next section to find the values of $\mathbf{A}$, $\mathbf{b}$, $\mathbf{c}$, $\bm{m}$, and~$d$ such that the resulting FDN is capable of modeling perceptually meaningful characteristics of the acoustic response of real-life environments. 

\section{Proposed Method}
\label{sec:proposed_method}

The proposed method involves an iterative gradient-based optimization algorithm. As a learning objective, we choose a perceptually-informed loss function, $\mathcal{L}(h, \hat{h})$, between a target RIR, $h[n]$, and the time-domain FDN output, $\hat{h}[n]$, obtained by setting the FDN input to the Kronecker delta, i.e., $u[n]=\delta[n]$. 

We initialize the FDN parameters with no prior knowledge of $h[n]$. Then, at the beginning of each iteration, we calculate $\hat{h}[n]$ by evaluating~\eqref{eq:time_domain_fdn} while freezing the current parameter estimates. Thus, we evaluate $\mathcal{L}(h, \hat{h})$. Finally, each trainable FDN parameter $\theta$ undergoes an optimization step using the error-free gradient $\nabla_\theta \mathcal{L}$ computed via reverse-mode automatic differentiation~\cite{baydin2018automatic}.

A typical approach is to use a delay network to only model the late reverberation while handling early reflections separately~\cite{bona2022avanzini, shen2020data, coggin2016automatic, ibnyahya2022method}. Instead, we optimize the FDN such that it accounts for both early and late reverberation at the same time, exploiting thus the advantages of synthesizing the entire RIR with an efficient recursive structure.

At training time, we strip out the initial silence due to direct-path propagation and disregard every sample beyond the $T_{60}$ of the target RIR. In other words, we only consider the first  $L_{T_{60}} := \left\lceil T_{60}\cdot f_\mathrm{s} \right\rceil$ samples of $h[n]$ and $\hat{h}[n]$ in computing the loss. The reason behind restricting the temporal scope only to the segment of the RIR associated with the $T_{60}$ is that, beyond this point, the values involved in the ensuing computations become so small that numerical errors might occur when using single-precision floating-point numbers, and the training process would unwantedly focus on the statistics of background/numerical noise. 
Notice that, at inference time, i.e., once the FDN parameters have been learned, the room acoustics simulation can be run indefinitely at a very low computational cost.

In this work, we optimize the input gains $\mathbf{b}\in\mathbb{R}_{\ge0}^N$, the output gains $\mathbf{c}\in\mathbb{R}_{\ge0}^N$, the direct gain $d\in\mathbb{R}_{\ge0}$, the feedback matrix $\mathbf{A}\in\mathbb{R}^{N\times N}$, and the %fractional
delays $\bm{m}\in\mathbb{R}_{\ge0}^N$ expressed in fractional samples.

\subsection{Model Reparameterization}
\label{ssec:reparameterization}
Let $\theta$ be a scalar parameter of the FDN such that $\theta\in\mathbb{X}$ where $\mathbb{X}\subseteq\mathbb{R}$. In general, instead of learning $\theta$ directly, we learn an unconstrained proxy $\tilde{\theta}\in\mathbb{R}$ that maps onto $\theta$ through a differentiable (and possibly nonlinear) function $f: \mathbb{R}\rightarrow\mathbb{X}$. 
Hence, we can use $f(\tilde{\theta})$ in place of $\theta$ in any computation involved in the forward pass of the FDN. 
In case of vector-valued parameters $\bm{\theta}\in\mathbb{X}^N$, we apply $f$ in an element-wise fashion, i.e., $\bm{\theta} := [f(\tilde{\theta}_1), ..., f(\tilde{\theta}_N)]^T$.
 
The reason behind such an explicit reparameterization method is that, while we would like $\bm{\theta}$ to take values in $\mathbb{X}^N$ at every iteration, gradient-based optimization may yield parameters that do not respect such a constraint, even when using implicit regularization strategies, e.g., by means of auxiliary loss functions and regularizers. 

In our FDN model, we treat every parameter in a different fashion. We discuss gain reparameterization in Section~\ref{ssec:trainable_gains} and present the feedback matrix reparameterization in Section~\ref{ssec:trainable_A}. Finally, we outline the implementation of the differentiable delay lines and their reparameterization in Section~\ref{ssec:diff_delay_lines}.

\begin{figure}[t]
    \centering
    \includegraphics[width=\linewidth]{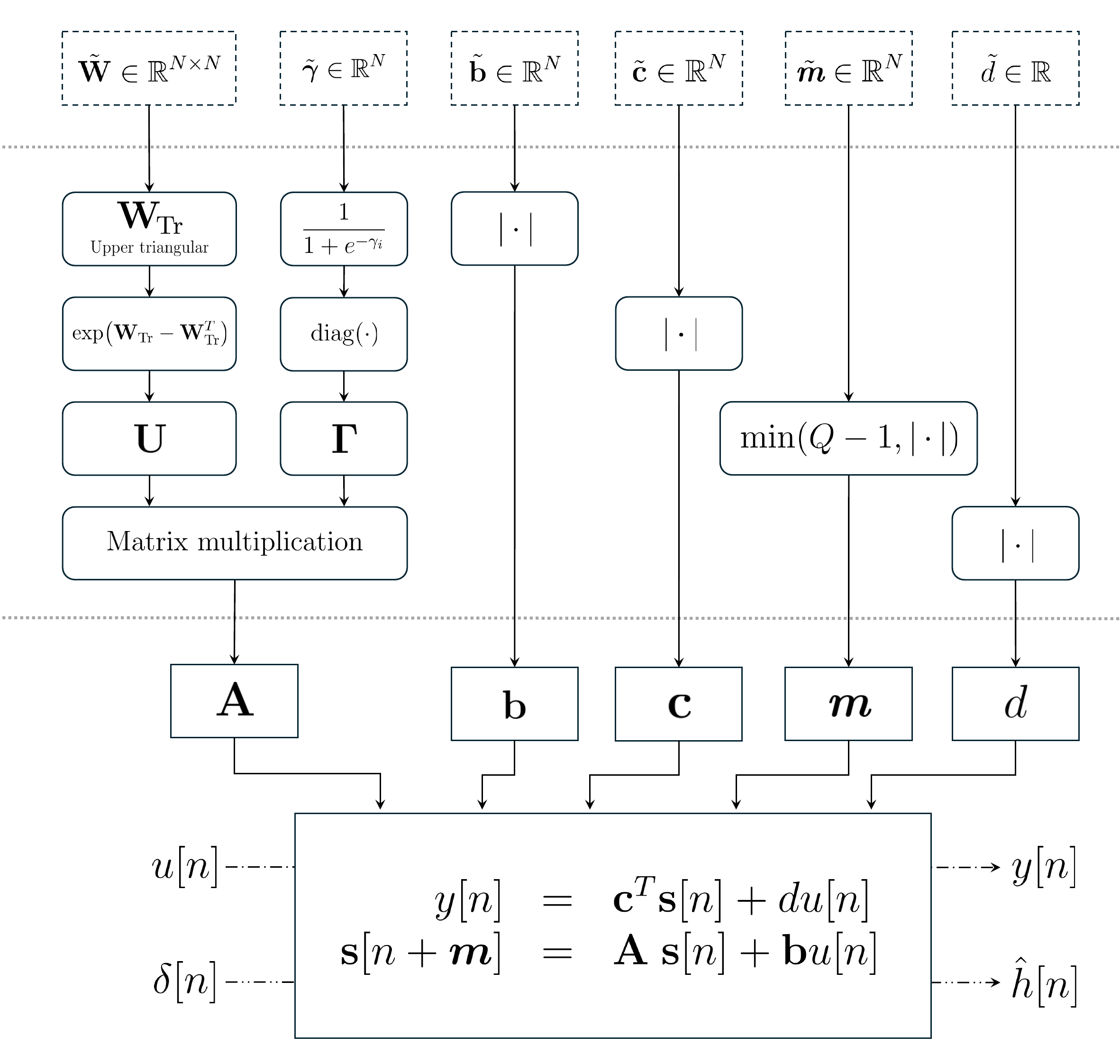}
    \caption{{\color{black} Summary of the proposed method.}}
    \label{fig:fdn_summary}
\end{figure}

{\color{black} 
Figure \ref{fig:fdn_summary} summarizes the proposed approach, listing all the unconstrained trainable parameters (top), the corresponding reparameterization (middle), and illustrating once again how, through \eqref{eq:time_domain_fdn}, the FDN processes time-domain signals, including unit impulses $\delta[n]$ (bottom).
}

\subsection{Trainable Gains}
\label{ssec:trainable_gains}
We would like the input, output, and direct gains of our differentiable FDN to be nonnegative. This way, the gains only affect the amplitude of the signals and do not risk inverting their polarity. Instead, we let $\mathbf{A}$ model phase-reversing reflections.
To enforce gain nonnegativity, we employ a differentiable nonlinear function $f_{\ge0}:\mathbb{R}\rightarrow\mathbb{R}_{\ge0}$, such as the Softplus or exponential function. We then learn, e.g., $\mathbf{\tilde{b}}=[\tilde{b}_1, ..., \tilde{b}_N]^T$ while using $\mathbf{b}=[f_{\ge0}(\tilde{b}_1), ..., f_{\ge0}(\tilde{b}_N)]^T$ in every computation concerning the FDN.
Among other options, we select $f_{\ge0}(x)= \lvert x\rvert$ \cite{mezza2023lumped}, where the requirement of $f_{\ge0}$ being differentiable everywhere was relaxed as it is common for many widely-adopted activation functions, such as ReLU.

\subsection{Trainable Feedback Matrix}
\label{ssec:trainable_A}
We focus on lossy FDNs. In prior work \cite{dalsanto2023colorless}, frequency-independent homogeneous decay has been modeled by parameterizing $\mathbf{A}$ as the product of a unilossless matrix $\mathbf{U}$ and a diagonal matrix 
$\bm{\Gamma}(\bm{m}) = \operatorname{diag}\left(\gamma_1, ..., \gamma_N\right) = \operatorname{diag}\left(\gamma^{m_1}, ..., \gamma^{m_N}\right)$
 containing a delay-dependent absorption coefficient for each delay line, where $\gamma\in(0,1)$ is a constant gain-per-sample parameter. The feedback matrix is thus expressed as
\begin{equation}
    \mathbf{A} = \mathbf{U}\bm{\Gamma}(\bm{m}).
    \label{eq:A_dal_santo}
\end{equation}

We let $\mathbf{U}$ be an orthogonal matrix, satisfying the unitary condition for unilosslessness~\cite{schlecht2016lossless}. To ensure this property, $\mathbf{U}$ is further parameterized by means of $\mathbf{\tilde{W}}\in\mathbb{R}^{N\times N}$ that, at each iteration, yields~\cite{dalsanto2023colorless}
\begin{equation}
    \mathbf{U} = \exp\!\left(\mathbf{W}_\text{Tr} - \mathbf{W}_\text{Tr}^T\right),
    \label{eq:unilossless_mapping}
\end{equation}
where $\mathbf{W}_\text{Tr}$ is the upper triangular part of $\mathbf{\tilde{W}}$, and $\exp(\cdot)$
is the matrix exponential.
In other words, instead of trying to directly learn a unilossless matrix, we learn an unconstrained real-valued matrix $\mathbf{\tilde{W}}$ that maps onto an orthogonal matrix through the exponential mapping in~\eqref{eq:unilossless_mapping}.\footnote{It is worth noting that, although $\mathbf{\tilde{W}}$ is a $N\times N$ matrix, only the $N(N-1)/2$ upper triangular entries are actually learned and used in downstream computations.}
In particular, $\mathbf{U}$ is ensured to be orthogonal because $\mathbf{W}_\text{Tr} - \mathbf{W}_\text{Tr}^T$ is skew-symmetric~\cite{lezcano2019cheap}.

As for the matrix $\bm{\Gamma}(\bm{m})$, we noticed that tying the values of the absorption coefficients $\gamma_1, ..., \gamma_N$ to those of the fractional delays $m_1, ..., m_N$ {\color{black} as previously done in \cite{dalsanto2023colorless}} led to instability during training since the values in $\bm{m}$ were concurrently acting on the temporal location of the IR taps as well as their amplitude.\footnote{This problem is unique to our approach, as previous studies employed non-trainable delay lines with fixed lengths~\cite{dalsanto2023colorless}.} 
Conversely, we decouple $\bm{\Gamma}$ from $\bm{m}$, thus learning a possibly inhomogeneous FDN characterized by $\mathbf{A} = \mathbf{U}\bm{\Gamma}$, as opposed to the homogeneous FDNs studied in~\cite{dalsanto2023colorless}.

In learning the unconstrained absorption matrix $\bm{\tilde{\Gamma}}=\operatorname{diag}\left(\tilde{\bm{\gamma}}\right)=\operatorname{diag}(\tilde{\gamma}_1, ..., \tilde{\gamma}_N)$, we define $f_{(0,1)}: \mathbb{R} \rightarrow (0,1)$ and optimize $\tilde{\bm{\gamma}}\in\mathbb{R}^N$ so that $\bm{\Gamma}=\operatorname{diag}\left(f_{(0,1)}(\tilde{\gamma}_1), ..., f_{(0,1)}(\tilde{\gamma}_N)\right)$.
In the following, we use the well-known Sigmoid function to force the absorption coefficients to take values in the range of $0$ to $1$, i.e., 
\begin{equation}
    f_{(0,1)}(x) = \frac{1}{1+e^{-x}}.
\end{equation}

\subsection{Trainable Delay Lines}
\label{ssec:diff_delay_lines}
In the digital domain, an integer delay can be efficiently implemented as a reading operation from a buffer that accumulates past samples. Unfortunately, this approach is not differentiable.
Instead, since the Fourier transform is a linear and differentiable operator, we opt to work in the frequency domain to circumvent the problem. 
%In particular, we evaluate the delays on the unit circle at discrete frequency points $\omega_k = 2\pi k/K$, for $k=0, ..., K-1$.

{\color{black}%
In~\cite{pei2012fractional}, Pei and Lai proposed a closed-form variable fractional delay filter, which turns out to be inherently differentiable.
In our FDN implementation, each delay line is equipped with a $Q$-sample buffer, so that the $i$th buffer stores the signal $x_i[n]$. 
First, we zero-pad $x_i[n]$ to reduce artifacts due to the ensuing circular convolution. 
Then, we compute the $K$-point Fast Fourier Transform (FFT) of the resulting signal, with $K=2Q$. Following~\cite{pei2012fractional}, we apply a delay of $m_i$ (fractional) samples by multiplying the discrete spectrum with the conjugate symmetric frequency response $D_i[k]$ defined in \eqref{eq:frac_delay_filter}. Finally, we go back in the time domain by computing the inverse FFT. We can express this sequence of differentiable operations as 
\begin{equation}
    x_i[n - m_i] = \operatorname{IFFT}\left\{ D_i[k] \cdot \operatorname{FFT}\left\{x_i[n]\right\}\right\},
    \label{eq:revised_delay_freq}
\end{equation}
where
\begin{equation}
    D_i[k] = \begin{cases}
        1, &k=0\\
        e^{-j m_i (2\pi/K)k}, &k=1,...,\frac{K}{2}-1\\
        \cos(m_i \pi), &k=\frac{K}{2}\\
        e^{-j m_i (2\pi/K)(K-k)}, &k=\frac{K}{2}+1, ..., K -1\\
    \end{cases}
    \label{eq:frac_delay_filter}
\end{equation}
which, in the time domain, corresponds to a windowed-sinc finite impulse response~\cite{pei2012fractional}.
%\begin{equation}
%h_{D_i}[n] = \frac{\cos\!\left(\frac{\pi(n-m_i)}{K}\right)}{\operatorname{sinc}\!\left(\frac{n-m_i}{K}\right)} \cdot \operatorname{sinc}(n-m_i)
%\end{equation}
%for $n=0, ..., K-1$.

Delays $m_1, ..., m_N$ must be nonnegative to realize a casual system. Hence, we use $f_{\ge0}$ to reparameterize them.
% to ensure that the fractional delays are nonnegative.
%i.e., $m_i = f_{\ge0}(\tilde{m}_i)$ where $\tilde{m}_i\in\mathbb{R}$ is the $i$th trainable delay-line length proxy, $i=1, ..., N$.
%Notice that such a parameterization, while enforcing nonegativity, imposes no upper bound on $m_i$. 
Moreover, we clip the resulting values so not to exceed the given buffer length.
This yields\footnote{{\color{black}
It is worth pointing out that \eqref{eq:delay_parametriz} is not the only way to account for the implicit periodization of the buffered signal when computing the FFT. For instance, an alternative parameterization is $m_i = (Q-1) \cdot f_{(0, 1)}(\tilde{m}_i)$, which ensures that $m_i\in(0, Q-1)$ at all times.}
}
\begin{equation}
    m_i = \min\left(Q-1, f_{\ge0}(\tilde{m}_i)\right),
    \label{eq:delay_parametriz}
\end{equation}
where $\tilde{m}_i\in\mathbb{R}$ is the $i$th trainable delay-line length proxy, $i=1, ..., N$.
%In our implementation, we set $Q=2^{13}$ (approximately $0.5$~s at a sampling rate of $16$ kHz) and $K=2Q$.
}

Finally, it is worth highlighting three main reasons for labeling our FDN model as \textit{time-domain}, despite implementing the differentiable delay lines in the frequency domain. 
First, we stress that our FDN yields the output one sample at a time according to~\eqref{eq:time_domain_fdn}. Second, frequency-domain operations are confined within the delay filterbank. {\color{black} Since delay lines being differentiable is only required at training time, the inference model can thus feature a different fractional delay implementation, possibly in the time domain.} Third, we emphasize the difference between our approach and existing methods implementing every FDN operation in the frequency domain~\cite{lee2022differentiable, dalsanto2023colorless}.

\subsection{Loss Function}
\label{ssec:loss}
Our goal is to learn an FDN capable of capturing perceptual qualities of a target room. Hence, we avoid pointwise regression objectives such as $L^p$-losses between IR taps. 
Instead, we set out to minimize an error function ($\mathcal{L}_\text{EDC}$) between the true and predicted EDCs. Additionally, we use a novel regularization loss ($\mathcal{L}_\text{EDP}$) aimed at matching the echo distribution of the target RIR by acting on the normalized EDP.
Namely, the composite loss function can be written as 
\begin{equation}
    \mathcal{L} = \mathcal{L}_\text{EDC} + \lambda\mathcal{L}_\text{EDP},
    \label{eq:tot_loss}
\end{equation}
where $\lambda\in\mathbb{R}_{\ge0}$. 
Similarly to \cite{mezza2023lumped}, the loss is evaluated in the time domain, and, at each iteration, requires a forward pass through the discrete-time model defined by the current parameter estimates. 
In the following sections, we analyze each of the terms in~\eqref{eq:tot_loss}.

\subsubsection{Energy Decay Curve Loss}
\label{sssec:edc_loss}
For a discrete-time RIR of length $L$, the EDC can be computed through Schroeder's backward integration~\cite{schroeder1965new}
\begin{equation} 
   \varepsilon[n] = \sum_{\tau=n}^L h^2[\tau].
   \label{eq:edc}
\end{equation}
Since \eqref{eq:edc} is differentiable, we can train the FDN to minimize a normalized mean squared error (NMSE) loss defined on the EDCs, i.e.,
\begin{equation}
    \mathcal{L}_\text{EDC} = \frac{\sum_n \left(\varepsilon[n] - \hat{\varepsilon}[n]\right)^2}{\sum_n \varepsilon[n]^2},
    \label{eq:edc_loss}
\end{equation}
where $\hat{\varepsilon}[n] = \sum_{\tau=n}^L \hat{h}^2[\tau]$.

It is worth noting that, whereas the EDC is typically expressed in dB, \eqref{eq:edc_loss} is evaluated on a linear scale. 
The idea here is that a linear loss emphasizes errors in the early portion of $\varepsilon[n]$, 
i.e., where discrepancies are perceptually more relevant~\cite{howard2013acoustics},
compared to a logarithmic loss that would put more focus on the reverberation tail. 

\subsubsection{Differentiable Normalized Echo Density Profile}
\label{sssec:edp_loss}

In~\cite{abel2006simple}, Abel and Huang introduced the so-called normalized Echo Density Profile (EDP) as a means to quantify reverberation echo density by analyzing consecutive frames of the reverberation impulse response. The EDP indicates the proportion of IR taps that fall above the local standard deviation. The resulting profile is normalized to a scale ranging from nearly zero, indicating a minimal presence of echoes, to around one, denoting a fully dense reverberation with Gaussian statistics~\cite{huang2007aspects}.

The EDP is defined as \cite{abel2006simple}
\begin{equation}
    \eta[n] =\frac{1}{\operatorname{erfc}(1/\sqrt{2})} \sum_{\tau=n-\nu}^{n+\nu} w[\tau]\mathds{1}\!\left\{ \lvert h[\tau]\rvert > \sigma_n \right\},
    \label{eq:edp}
\end{equation}  
where $\operatorname{erfc}(\cdot)$ is the complementary error function,
\begin{equation}
    \sigma_n = \sqrt{\sum_{\tau=n-\nu}^{n+\nu} w[\tau]h^2[\tau]},
\end{equation}
is the standard deviation of the $n$th frame, $w[n]$ is a window function of length $2\nu + 1$ samples (usually $20$~ms) such that $\sum_\tau w[\tau] = 1$, and $\mathds{1}\{\cdot\}$ is an indicator function
\begin{equation}
    \mathds{1}\!\left\{ \lvert h[\tau]\rvert > \sigma \right\} = \left\{
    \begin{array}{cc}
        1 &\quad \lvert h[\tau]\rvert > \sigma,\\
        0 &\quad \lvert h[\tau]\rvert \le \sigma.
    \end{array}
    \right.
\end{equation}
Notably, $\mathds{1}\{\cdot\}$ is non-differentiable. Thus, the EDP cannot be utilized within our automatic differentiation framework.

To overcome this problem, this section introduces a novel differentiable EDP approximation, which we call \textit{Soft Echo Density Profile}. 

First, we notice that the indicator function $\mathds{1}\!\left\{ \lvert h[\tau]\rvert > \sigma \right\}$ can be equivalently expressed as a Heaviside step function $ \mathcal{H}\left(\lvert h[\tau]\rvert - \sigma \right)$.
Then, we let $g(x)$ denote the Sigmoid function. 
We define the \textit{scaled Sigmoid} function $g_\kappa(x) = g(\kappa x)$, where $\kappa\in\mathbb{R}_{>0}$. 
Since
\begin{equation}
    \lim_{\kappa\rightarrow\infty} g_\kappa(x) = \mathcal{H}(x),
\end{equation}
we can define the Soft EDP function as
\begin{equation}
    \eta_{\kappa}[n] =\frac{1}{\operatorname{erfc}(1/\sqrt{2})} \sum_{\tau=n-\nu}^{n+\nu} w[\tau]g_{\kappa}\! \left( \lvert h[\tau]\rvert - \sigma_n \right), 
    \label{eq:soft_edp}
\end{equation}
which approximates \eqref{eq:edp} for $\kappa\gg1$.

It is worth mentioning that, whilst the EDP approximation  improves as $\kappa$ becomes larger, this also has the side effect of increasing the risk of vanishing gradients. In fact, the derivative of the scaled Sigmoid function can be written as
\begin{equation}
    g'_\kappa(x) = g(\kappa x)\left( 1 - g(\kappa x)\right),
\end{equation}
which approaches zero for large or small inputs. Hence, $g'_\kappa(x)$ takes on near-zero values outside of a neighborhood of $x=0$ whose size is inversely proportional to $\kappa$, which, in turn, may impede the gradient flow for $\kappa\gg1$.

In practice, we would like to choose a large value of $\kappa$ but not larger than what is needed. Notably, the need for a large scaling factor is not constant throughout the temporal evolution of a RIR. Early taps are typically sparse, and $\left(\lvert h[\tau]\rvert - \sigma_n\right)$ tends to fall within the saturating region of $g_\kappa(\cdot)$, even for lower values of $\kappa$. Conversely, in later portions of the RIR, $\kappa$ must take on very large values to contrast the fact that the amplitude of $\left(\lvert h[\tau]\rvert - \sigma_n\right)$ progressively decreases. For this reason, we introduce a time-varying scaling parameter, $\kappa_n=\xi n + \varrho$, where $\xi\in\mathbb{R}_{> 0}$ and $\varrho\in\mathbb{R}_{\ge 0}$ are hyperparameters. Progressively increasing the scaling coefficient has the benefit of enhancing the gradient flow for the early reflections, while improving the EDP approximation for late reverberation. In general, a more principled definition for $\kappa_n$ could be devised, e.g., by tying it to the local statistics of the target RIR or its energy decay. In this work, however, we favor a simple and reproducible approach as it proved to work well in practice.

\subsubsection{Soft EDP Loss}
Despite the trade-off between vanishing gradients and goodness of fit discussed in the previous section, every operation involved in the computation of \eqref{eq:soft_edp} is   differentiable {\color{black} almost everywhere}. This allows us to use the following EDP loss term as a regularizer during the FDN training
\begin{equation}
    \mathcal{L}_\text{EDP} = \frac{1}{L_{T_{60}}}\sum_n \left(\eta_\kappa[n] - \hat{\eta}_\kappa[n]\right)^2,
    \label{eq:edp_loss}
\end{equation}
where $\hat{\eta}_\kappa[n]$ is the Soft EDP of the predicted RIR, and $L_{T_{60}}=\left\lceil T_{60}\cdot f_\mathrm{s} \right\rceil$.

\section{Evaluation}
\label{sec:evaluation}
We evaluate the proposed method using real-world measured RIRs from the 2016 MIT Acoustical Reverberation Scene Statistics Survey \cite{mit2016}.
The MIT corpus contains single-channel environmental IRs of both open and closed spaces.
Of the $271$ IRs, we select three according to their reverberation time, which, across the dataset, ranges from a minimum of $0.06$~s to a maximum of $1.99$~s. We select three indoor environments:%
\footnote{{\color{black} Although the proposed parameter tuning method shares some similarities with neural network training, particularly in their use of backpropagation, differentiable FDNs require a dedicated optimization routine for each target RIR. When it comes to evaluation, this study thus focuses on a limited number of illustrative examples; this approach is consistent with white-box system identification literature while contrasting with the way deep learning models are typically evaluated, which, instead, involves large-scale training and test sets.
}}
(i) a small room ($T_{60}\approx0.2$~s), (ii) a medium room ($T_{60}\approx0.6$~s), and (iii) a larger room ($T_{60}\approx1.2$~s).
For reproducibility, the ID of the chosen RIRs is reported:
(i) \texttt{h214\_Pizzeria\_1txts}, 
(ii) \texttt{h270\_Hallway\_House\_1txts}, 
and (iii) \texttt{h052\_Gym\_WeightRoom\_3txts}.
The full IR Survey dataset is available online.\footnote{[Online] IR Survey dataset: \url{https://mcdermottlab.mit.edu/Reverb/IR_Survey.html}}
For the evaluation, all IRs are resampled to 16~kHz and scaled to unit norm.

\subsection{Baseline Methods}
From the overview presented in Section~\ref{sec:related_work}, it appears that no method in the literature is directly comparable with ours. In fact, existing automatic tuning approaches either focus on off-the-shelf reverb plug-ins~\cite{heise2009automatic, bona2022avanzini}, limit the set of target parameters to just a few~\cite{chemistruck2012generating, shen2020data}, or augment the FDN topology with auxiliary frequency-dependent components~\cite{coggin2016automatic, ibnyahya2022method}.
To the best of our knowledge, there is no state-of-the-art method addressing the simultaneous estimation of every parameter of a  time-invariant
frequency-independent FDN in a purely data-driven fashion. %with the objective of matching a target RIR.

That being said, with the aim of comparing our approach with existing techniques, we implement three baseline methods.

The first is based on a classic method for homogeneous reverberation time control~(HRTC) and involves choosing all FDN parameters but the absorption coefficients heuristically. For simplicity, we refer to this method as ``HRTC baseline.''

The second method, which we call ``Colorless baseline,'' also relies on HRTC to control the decay rate. However, contrary to the HRTC baseline, all remaining FDN parameters are optimized following the approach detailed in~\cite{dalsanto2023colorless} as so to achieve a maximally flat frequency response.

The third and final baseline, inspired by \cite{ibnyahya2022method}, makes use of a genetic algorithm~(GA) to optimize every FDN parameter except for the feedback matrix. We call this method ``GA baseline.''

In the following sections, we detail the three baseline methods one at a time.

\subsubsection{HRTC Baseline}
\label{sssec:hrtc_baseline}
Given the FDN model shown in Figure~\ref{fig:siso_fdn}, a classic method to introduce homogeneous loss in an otherwise lossless prototype is to replace each unit delay $z^{-1}$ with a lossy delay element $\gamma z^{-1}$, where $\gamma$ is thought of as a gain-per-sample coefficient \cite{jot_fdn, schlecht2020fdntb, dalsanto2023colorless}. 

In practice, the loss of each delay line is lumped into a single attenuation term proportional to its length. We can thus define $\bm{\Gamma}(\bm{m}) = \operatorname{diag}\left(\gamma^{m_1}, ..., \gamma^{m_N}\right)$ as discussed in Section \ref{ssec:trainable_A}, where $\gamma$ controls the decay rate according to the desired reverberation time. Namely, $\gamma$ should satisfy \cite{jot_fdn}
\begin{equation}
    20 \log_{10} \gamma = \frac{-60}{f_\mathrm{s} T_{60}},
    \label{eq:gamma_db_T60}
\end{equation}
where $T_{60}$ is estimated from the target RIR.

Whereas the absorption coefficients are given by \eqref{eq:gamma_db_T60}, every other parameter in the HRTC baseline are determined by means of heuristics. 
We parameterize the feedback matrix as in \eqref{eq:A_dal_santo}, where $\mathbf{U}$ is a random orthogonal matrix. To ensure that the HRTC baseline is most comparable with the proposed method (Section \ref{ssec:trainable_A}) and the Colorless baseline (Section \ref{sssec:colorless_baseline}), we obtain $\mathbf{U}$ through the exponential mapping in \eqref{eq:unilossless_mapping}.  
The scalar gain $d$ is set equal to the amplitude of the target RIR at the time index associated with the direct path.
We use unity input gains, i.e., $\mathbf{b}=\bm{1}_N$, where $\bm{1}_N$ is a vector of $N=6$ ones. The output gains are chosen so that $\mathbf{c}=\frac{1}{N}\bm{1}_N$.
As such, the dot product $\mathbf{c}^T\mathbf{s}[n]$ in \eqref{eq:time_domain_fdn} is equivalent to the arithmetic average of the outputs of the $N$ delay lines at time $n$. 
Finally, the delays $\bm{m}=[997, 1153, 1327, 1559, 1801, 2099]$ consist of logarithmically distributed prime numbers 
%in the range of $20$~ms to $43.7$~ms 
from Delay Set \#1 in \cite{dalsanto2023colorless}, and the corresponding non-differentiable integer delay lines are implemented via buffer readout.

\subsubsection{Colorless Baseline}
\label{sssec:colorless_baseline}

In the previous section, we discussed a baseline method consisting of a homogeneous FDN where the reverberation time is controlled by choosing $\gamma$ according to \eqref{eq:gamma_db_T60}, with the other parameters being manually selected. Here, we present an alternative baseline method that foregoes some of these arbitrary choices in favor of an optimization approach.

In \cite{dalsanto2023colorless}, the authors implement a differentiable homogeneous FDN in the frequency domain and find $\mathbf{A}$, $\mathbf{b}$, and $\mathbf{c}$ via gradient descent so as to minimize spectral coloration. 

Colorless reverberation \cite{schroeder1961colorless} is here defined as the acoustic quality of an artificial reverberation algorithm whose frequency response is flat, i.e., constant at all frequencies.

To achieve this, $\mathbf{A}$, $\mathbf{b}$, and $\mathbf{c}$ are iteratively updated via backpropagation using Adam \cite{Kingma2015adam} to minimize a reference-free\footnote{With \textit{reference-free}, we emphasize that, unlike our method, the loss function in \cite{dalsanto2023colorless} is computed solely on the FDN response, and does not consider a reference RIR as the target of the optimization process.} loss function comprising two terms \cite{dalsanto2023colorless}. The first term encourages the magnitude of the sampled transfer function of each delay network channel to be as close to one as possible. The second term penalizes IR sparsity in the time domain and avoids trivial solutions. 

The delays $\bm{m}$ are kept constant, and $\mathbf{A}$ is parameterized through \eqref{eq:A_dal_santo} and \eqref{eq:unilossless_mapping}. Like in the HRTC baseline, we use the lengths in samples comprised in Delay Set~\#1 from~\cite{dalsanto2023colorless}, and $\gamma$ is set according to~\eqref{eq:gamma_db_T60}.

With this baseline, our goal is to test whether an FDN optimized to obtain a flat magnitude response brings about any benefit when it comes to modeling the energy decay and echo density of a target RIR.
It is worth noting, however, that the learning objective in \cite{dalsanto2023colorless} is not concerned with matching the behavior of a reference RIR. As such, the resulting $\mathbf{A}$, $\mathbf{b}$, and $\mathbf{c}$ might prove suboptimal for what concerns reproducing the reverberation time of the target space.

\subsubsection{GA Baseline}
\label{sssec:ga_baseline}

\begin{figure}[t]
    \centering
    \includegraphics[width=\linewidth]{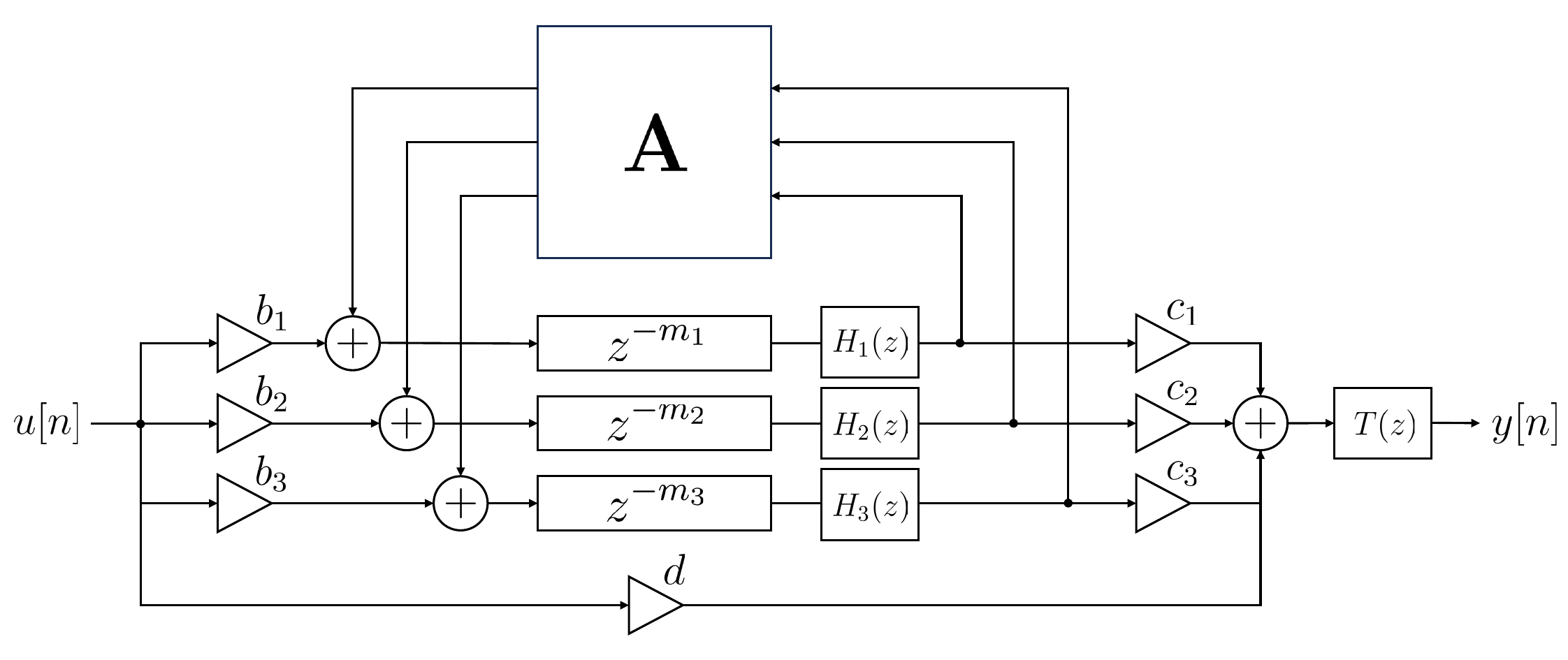}
    \caption{Block diagram of the prototype FDN used in \cite{ibnyahya2022method} with $N=3$. }
    \label{fig:baseline_fdn}
\end{figure}

In~\cite{ibnyahya2022method}, Ibnyahya and Reiss proposed a multi-stage automatic tuning approach that combines genetic optimization~\cite{goldberg1989genetic} and analytical filter design~\cite{schlecht2017accurate}. 
{\color{black}
Adhering to well-established design principles~\cite{jot_fdn}, the prototype FDN considered in~\cite{ibnyahya2022method} is equipped with attenuation filters $H_i(z)$ that modify the frequency and total energy of the normal modes of the system's response, and a tone-correction filter $T(z)$~\cite{valimaki2016accurate} that modify the system's power spectral density by imposing a desired magnitude frequency response. 
While we depict the model architecture with $N=3$ in Figure~\ref{fig:baseline_fdn}, our implementation uses $N=6$.
}

As in~\cite{ibnyahya2022method}, we aim to optimize $\bm{m}$, $\mathbf{b}$, $\mathbf{c}$, and $d$, whereas $\mathbf{A}$ is fixed throughout the procedure. 
The GA is run for $50$ generations (i.e., ten times more than in~\cite{ibnyahya2022method}), each with a population of $50$ FDNs. Each FDN is therefore an \textit{individual} characterized by $3N + 1$ mutable parameters, namely, $\bm{m}$, $\mathbf{b}$, $\mathbf{c}$, and $d$.
During the optimization, scalar gains are constrained to take values in $[-1, 1]$. Similarly, delays are constrained to take values in the range of 200~\textmu s to 64~ms.

The attenuation filters are analytically determined according to the individual's delay values in $\bm{m}$ and the desired octave-band reverberation times~\cite{schlecht2017accurate}. In turn, the output graphic EQ filter~\cite{valimaki2016allabout} is found based on the initial level of the desired octave-band EDCs.
All individuals implement the same random orthogonal feedback matrix~\cite{edelman_rao_2005}, which is not affected by genetic optimization. 
The fitness of each individual at every generation is assessed through the mean absolute error between the MFCCs of the target RIR and those of the FDN output \cite{ibnyahya2022method}.

In our implementation, we avail of the Feedback Delay Network Toolbox by S.\ J.\ Schlecht~\cite{schlecht2020fdntb} for fitting the graphic EQ filters and implementing the FDN, and use the GA solver included in MATLAB's Global Optimization Toolbox for finding $\bm{m}$, $\mathbf{b}$, $\mathbf{c}$, and $d$.

It is worth emphasizing the differences between the prototype FDN used in~\cite{ibnyahya2022method} (Figure~\ref{fig:baseline_fdn}) and the proposed delay network (Figure~\ref{fig:siso_fdn}).
First, \cite{ibnyahya2022method}~does not optimize the feedback matrix $\mathbf{A}$, whereas we do.
Second, \cite{ibnyahya2022method}~relies on IIR filters to achieve the desired reverberation time, whereas our model does not. Introducing $H_i(z)$ and $T(z)$ makes the baseline arguably more powerful in modeling a target RIR. At the same time, though, prior knowledge must be injected into the model by means of filter design to successfully run the GA and obtain meaningful results in a reasonable number of generations. 

\begin{table}[t]
\centering
\caption{Reverberation time, iteration indices, loss values, and average time per training step {\color{black} in seconds (NVIDIA Tesla V100)}. Iterations denoted with $(0)$ indicate pre-training random initialization.}\label{tab:loss_table}%
\setlength{\tabcolsep}{0pt}
\begin{tabular*}{\linewidth}{@{\extracolsep{\fill}}lccccccc@{}}             \toprule
     & $T_{60}$ & iter & $\mathcal{L}$ & $\mathcal{L}_\text{EDC}$ & $\mathcal{L}_\text{EDP}$ & time [s]\\
    \midrule
    \multirow{2}{*}{Gym} & \multirow{2}{*}{$1.225$}&  $(0)$ & $0.9959$ & $0.9768$ & $0.1909$ & \multirow{2}{*}{$41.86$} \\
      & & $935$ & $0.0067$ & $0.0058$ & $0.0095$ & &\\
    \midrule
    \multirow{2}{*}{Hallway} & \multirow{2}{*}{$0.607$} & $(0)$ & $1.0068$ & $0.9812$ & $0.2560$ & \multirow{2}{*}{$20.76$}\\
      & & $796$ &$0.0041$ & $0.0034$ & $0.0068$ & &\\
    \midrule
    \multirow{2}{*}{Pizzeria} & \multirow{2}{*}{$0.206$} & $(0)$ & $0.9254$ & $0.9038$ & $0.2161$ &  \multirow{2}{*}{$5.71$}\\
      & & $992$ & $0.0526$ & $0.0501$ & $0.0255$ & &\\
    \bottomrule
\end{tabular*}
\end{table}

\subsection{Evaluation Metrics}
As evaluation metrics, we select the $T_{20}$, $T_{30}$, and $T_{60}$, i.e., the reverberation time extrapolated considering the normalized IR energy decaying from $-5$~dB to $-25$~dB, $-35$~dB, and $-65$~dB, respectively. 
Ideally, these three metrics are the same if the EDC exhibits a perfectly linear slope. In practice, this is often not the case, as it can be seen, e.g., in Figure~\ref{fig:edc_gym}. 
Hence, we believe that it is more informative to report all three of them, as together they provide a richer insight into the global behavior of the EDC as it approaches the $-60$~dB threshold. 
It is also worth pointing out that it is unclear whether the $T_{60}$ is entirely reliable in measuring the reverberation time of real-world RIRs due to the often non-negligible noise floor.

Furthermore, we report the following ISO 3382 measures~\cite{iso3382}: \textit{Clarity}~($C_{80}$), expressed in dB, \textit{Definition}~($D_{50}$), expressed as a percentage, and \textit{Center time}~($t_s$), expressed in ms. 
Having defined $L_{\tau} := \left\lceil \tau \cdot f_\mathrm{s} \cdot 10^{-3} \right\rceil$, these metrics are given by
\begin{align}
    C_{80} = 10 \log_{10} \frac{\sum_{n=0}^{L_{80}-1}h^2[n]}{\sum_{n=L_{80}}^{L-1} h^2[n]}&,\\
    D_{50} = 100 \cdot \frac{\sum_{n=0}^{L_{50}-1}h^2[n]}{\sum_{n=0}^{L-1} h^2[n]}&,\\
    t_s = 10^{3} \cdot \frac{\sum_{n=0}^{L-1} n \cdot h^2[n]}{f_\mathrm{s}\cdot\sum_{n=0}^{L-1} h^2[n]}&.
\end{align}
For each metric, we report the error with respect to the target values. Absolute deviations are denoted by $\Delta$ in Tables~\ref{tab:metrics_gym}, \ref{tab:metrics_hallway}, and \ref{tab:metrics_pizzeria}.

Finally, it is worth pointing out that the EDPs shown in the following sections are obtained using the (non-differentiable) formulation given in \eqref{eq:edp} unless explicitly stated otherwise.

\subsection{Parameter Initialization}
\label{ssec:initialization}
We initialize the differentiable FDN parameters as follows. We let $\mathbf{\tilde{b}}^{(0)} \sim \mathcal{N}(\bm{0}, \frac{1}{N}\mathbf{I}_N)$, where $\mathbf{I}_N$ is the $N\times N$ identity matrix. We let $\mathbf{\tilde{c}}^{(0)}=\frac{1}{N}\bm{1}_N$, where $\bm{1}_N$ is a vector of $N$ ones. We set $\tilde{d}^{(0)}=1$.
We initialize $\mathbf{\tilde{W}}^{(0)}$ and $\bm{\tilde{\Gamma}}^{(0)}$ so that $\mathbf{\tilde{W}}^{(0)}_{ij}\sim\mathcal{N}(0, \frac{1}{N})$ and $\tilde{\gamma}^{(0)}_{i}\sim\mathcal{N}(0, \frac{1}{N})$.
We initialize $\tilde{\bm{m}}^{(0)}$ so that $\tilde{m}^{(0)}_i = \psi \tilde{m}_i^\star$ with $\tilde{m}_i^\star \sim \text{Beta}(\alpha, \beta)$, for $i=1, ..., N$, where $\alpha \ge 1$ and $\beta>\alpha$. We empirically set $\psi=1024$, $\alpha=1.1$, and $\beta=6$ to ensure a maximum possible delay of $64$~ms (the same as in the GA baseline) and a mean value of about $10$~ms. 
We let the Sigmoid scaling term $\kappa_n$ increase linearly from $10^2$ to $10^5$ as $n=0, ..., L_{T_{60}}-1$. 
% Finally, we choose the hyperparameter $\lambda$ for each test case (see Table~\ref{tab:loss_table}) in order to balance the two terms in~\eqref{eq:tot_loss}.

\subsection{Implementation Details}
We implement our differentiable model in Python using PyTorch. We define the FDN as a class inheriting from \texttt{nn.Module}. We thus define the unconstrained trainable parameters as instances of \texttt{nn.Parameter}.
Our model operates at a sampling rate of 16~kHz.
As a result, its memory footprint turns out to be contained, allowing us to train all FDNs considered in the present study on a single 16~GB NVIDIA Tesla V100 graphics card.\footnote{Preliminary experiments carried out with increased computational resources indicate that results comparable with those reported in the present study can be obtained at a sampling rate of 48~kHz.}
We optimize the models for a maximum of $1000$ iterations using Adam~\cite{Kingma2015adam} with a learning rate of $0.1$, $\beta_1=0.9$, $\beta_2=0.999$, and no weight decay. 
{\color{black} 
In all test cases, the EDP loss term is weighted by  $\lambda=0.1$.
}

The average training time per iteration is reported in Table \ref{tab:loss_table}. At each step, just below 24\% of the time is taken by the forward pass of the FDN, approximately 12\% is spent computing the loss function, 
and just above 64\% is spent backpropagating the gradients and updating the parameters.
Notably, we observe that the computation time increases linearly with the estimated $T_{60}$.
For each test case, we present the model with the lowest composite loss. 

\subsection{Test Case: Gym  (h052)}
\label{ssec:test_case_gym}
\begin{figure}[t]
    \centering
    \includegraphics[width=\linewidth]{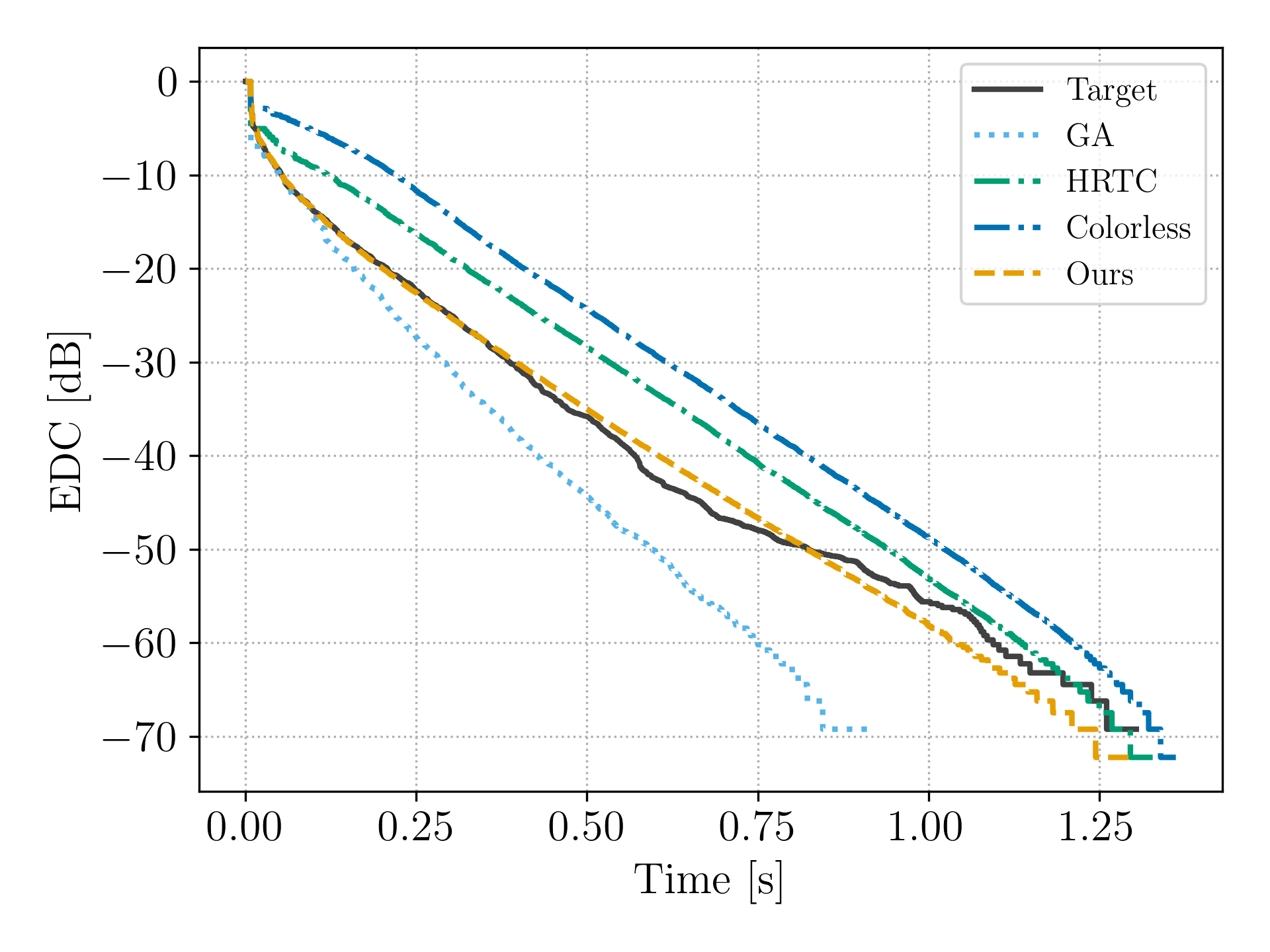}
    \vspace{-1em}
    \caption{Gym (\texttt{h052}) EDCs.}
    \label{fig:edc_gym}
\end{figure}

\begin{figure}[t]
    \centering
    \includegraphics[width=\linewidth]{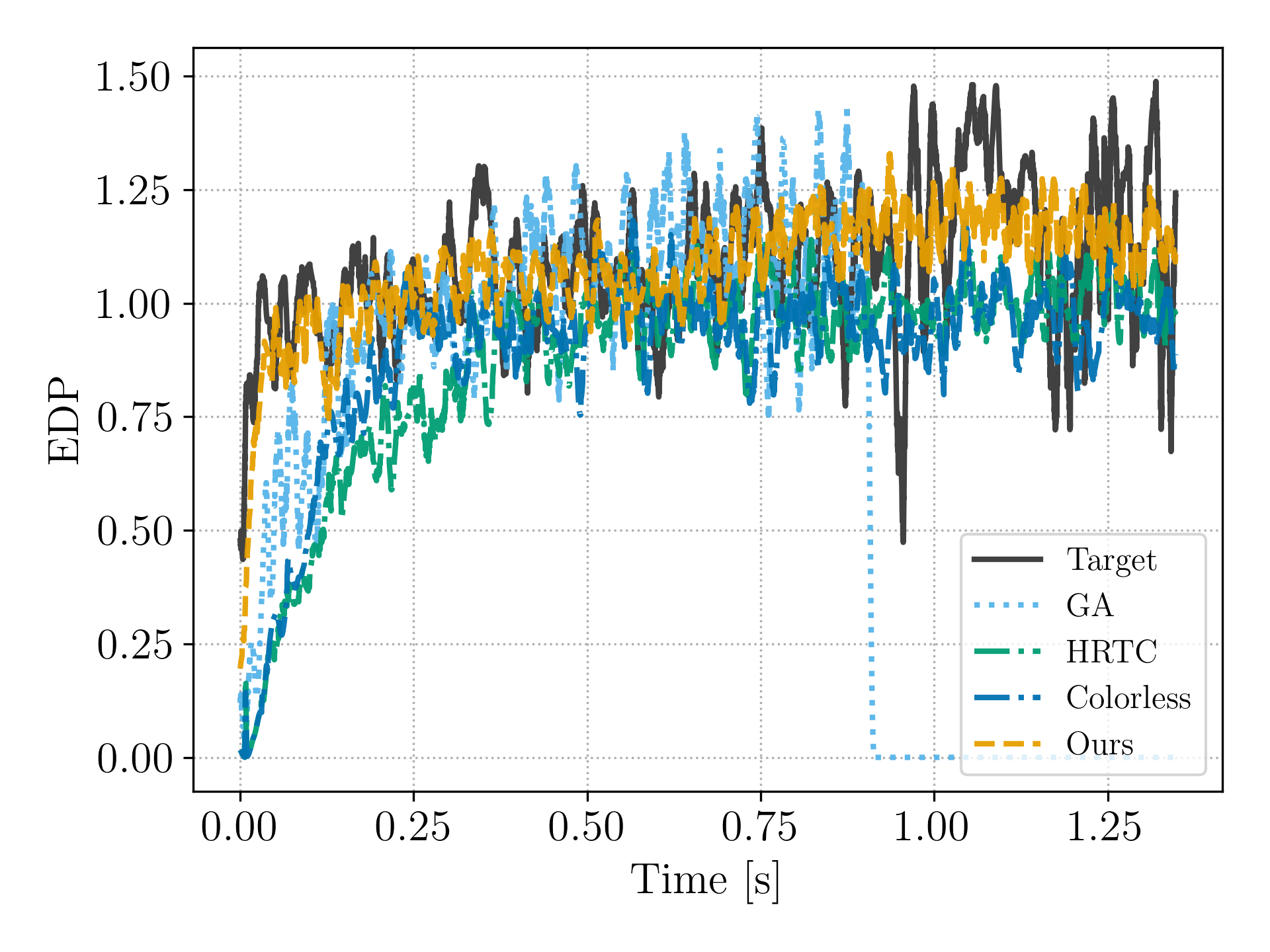}
    \vspace{-1em}
    \caption{Gym (\texttt{h052}) EDPs.}
    \label{fig:edp_gym}
\end{figure}

\begin{figure}[t]
    \centering
    \includegraphics[width=\linewidth]{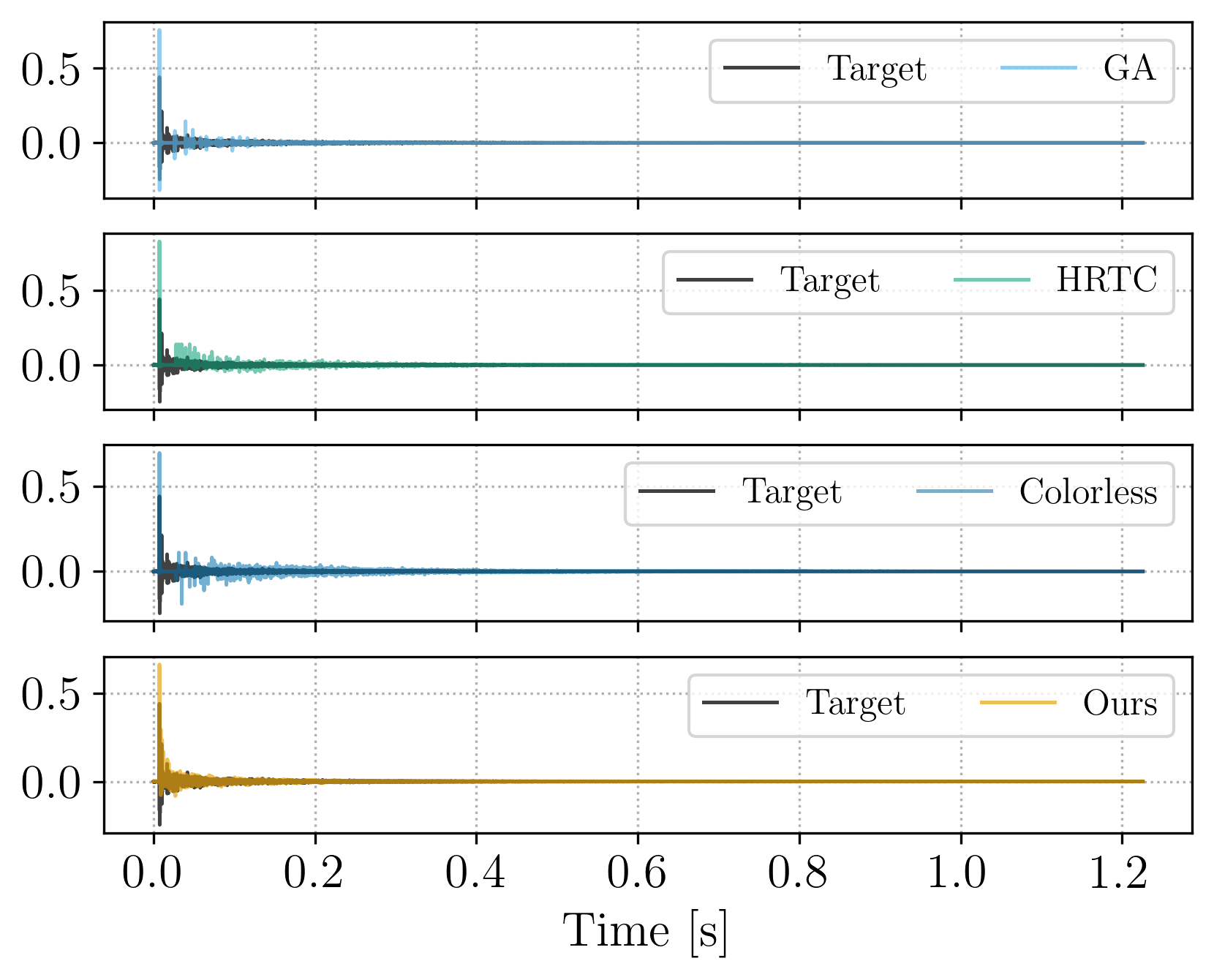}
    \vspace{-1em}
    \caption{Gym (\texttt{h052}) IRs. {\color{black} The time axis is limited to the $T_{60}$ for visual clarity.}}
    \label{fig:rir_gym}
\end{figure}

{\color{black}
We start by considering the Gym RIR (\texttt{h052}). 
%The FDN is trained setting $\lambda=1$, i.e., equally weighting $\mathcal{L}_\text{EDC}$ and $\mathcal{L}_\text{EDP}$. 
As shown in Table~\ref{tab:loss_table}, the best model is reached at iteration 935, after the loss has decreased by three orders of magnitude with respect to the initial value obtained with the random initialization described in Section~\ref{ssec:initialization}.
Here, we present a comparison between the target room acoustics (solid black line), the GA baseline~\cite{ibnyahya2022method} (dotted blue line), the HRTC baseline (dash-dotted green line), the Colorless baseline~\cite{dalsanto2023colorless} (dash-dotted blue line), and, finally, the proposed differentiable FDN (dashed orange line). 

Figure~\ref{fig:edc_gym}, \ref{fig:edp_gym}, and \ref{fig:rir_gym} show that the proposed method is capable of closely matching the EDC, EDP, and envelope of the target RIR, respectively.
Conversely, the baseline methods produce poorer results. 

In Figure~\ref{fig:edc_gym}, we may notice that the EDC of the GA baseline deviates from that of the target RIR after just 100~ms and exhibits an overall steeper decay. The EDC of both the HRTC and Colorless baselines, instead, overshoot the target after the very first few ms before decaying with a linear slope. It is interesting to notice, though, that HRTC approaches and matches the target EDC at around the estimated $T_{60}$, i.e., $1.225$~s, differently from the Colorless baseline, which had overshoot the target curve more. 
%Such a difference could be due to the feedback matrix and input/output gains of the latter being optimized to get a colorless dense IR, which, in turn, may have affected the correct time decay rendering.

In Figure~\ref{fig:edp_gym}, the EDP of the baseline methods indicate a scarce echo density in the first 250~ms, especially for the HRTC and Colorless methods. The output of the FDN obtained with GA, instead, becomes identically zero just after approximately 0.9~s. The ensuing EDP pathologies are confirmed by the IR depicted in Figure~\ref{fig:rir_gym}, where the baselines are shown to yield fewer, more prominent taps compared to the IR of the proposed FDN model. % shown in Figure~\ref{fig:rir_gym}.

Notably, Figure~\ref{fig:rir_gym} also suggests that all methods estimate a larger $d$ than what would correctly render the direct sound. We attribute this phenomenon to an attempt at compensating the lack of a noise floor that, in real-life measurements, contributes to the total energy of the RIR. We argue that, in FDN models with tunable direct gain, offsetting this bias is naively achieved by increasing $d$. 

Table~\ref{tab:metrics_gym} shows that the proposed method has an overall better performance in five out of the six reverberation metrics, with a $\Delta T_{20}$, $\Delta T_{30}$, $\Delta C_{80}$, $\Delta D_{50}$, and $\Delta t_s$ of 16.5~ms, 55.2~ms, 0.02 dB, 0.09\%, and 180~\textmu s, respectively. In particular, $T_{20}$, $T_{30}$, and $t_s$ are estimated with an error over one order of magnitude lower than those of the other methods. Likewise, the proposed FDN improves upon the baselines by two orders of magnitude as far as clarity $C_{80}$ and definition $D_{50}$ are concerned. On the contrary, the proposed approach yields and error of 90.2~ms when it comes to the $T_{60}$, and it is surpassed by HRTC and Colorless, whose $\Delta T_{60}$ is 6.3~ms and 35.7~ms, respectively. This, however, was largely expected given that, in both baseline methods, the parameter $\gamma$ is specifically designed to match the desired $T_{60}$ according to \eqref{eq:gamma_db_T60}.
}

% -------------------------------------------------


% -------------------------------------------------

\begin{figure}[t]
    \centering
    \includegraphics[width=\linewidth]{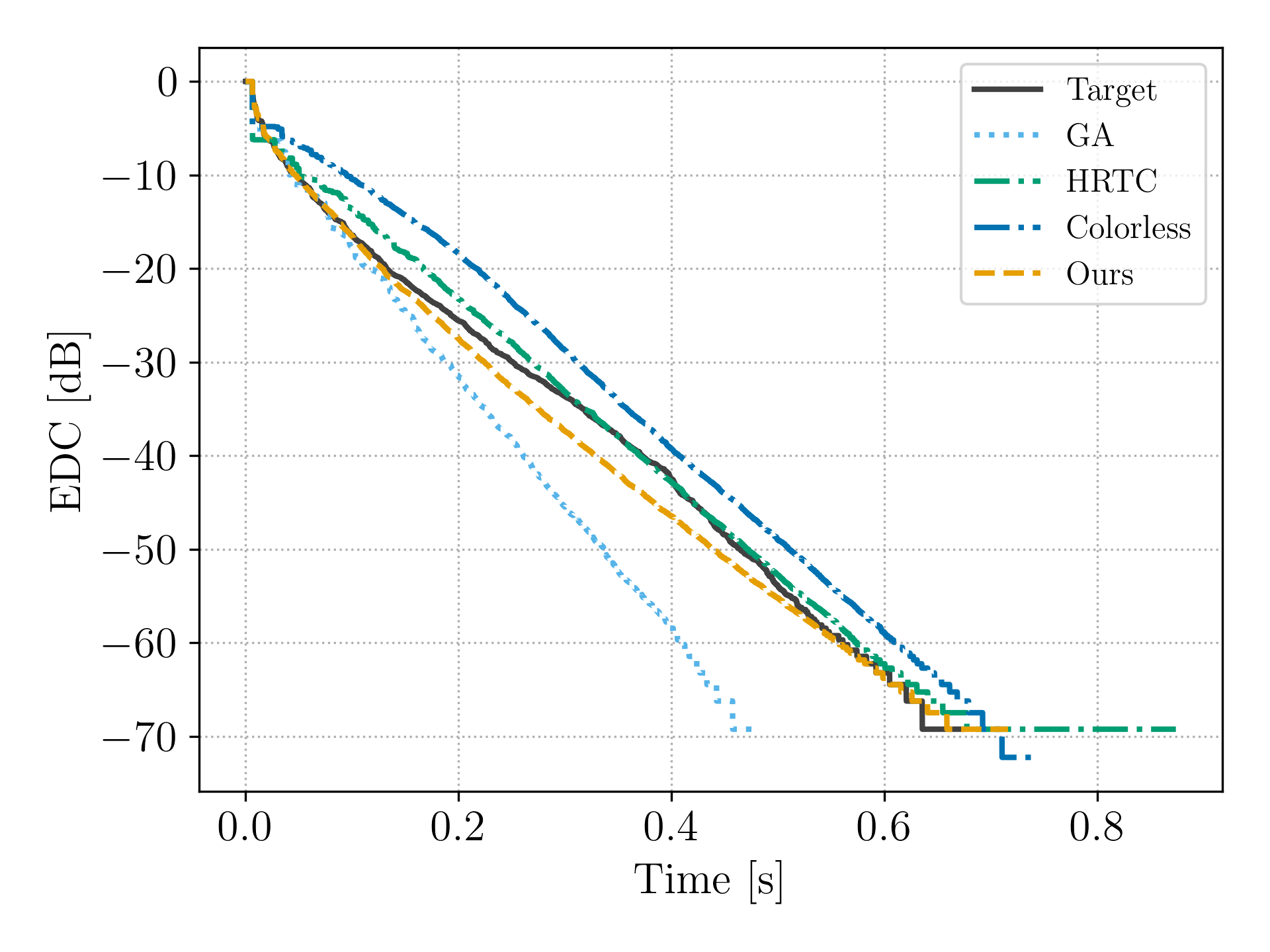}
    \vspace{-1em}
    \caption{Hallway (\texttt{h270}) EDCs.}
    \label{fig:edc_hallway}
\end{figure}
\begin{figure}[t]
    \centering
    \includegraphics[width=\linewidth]{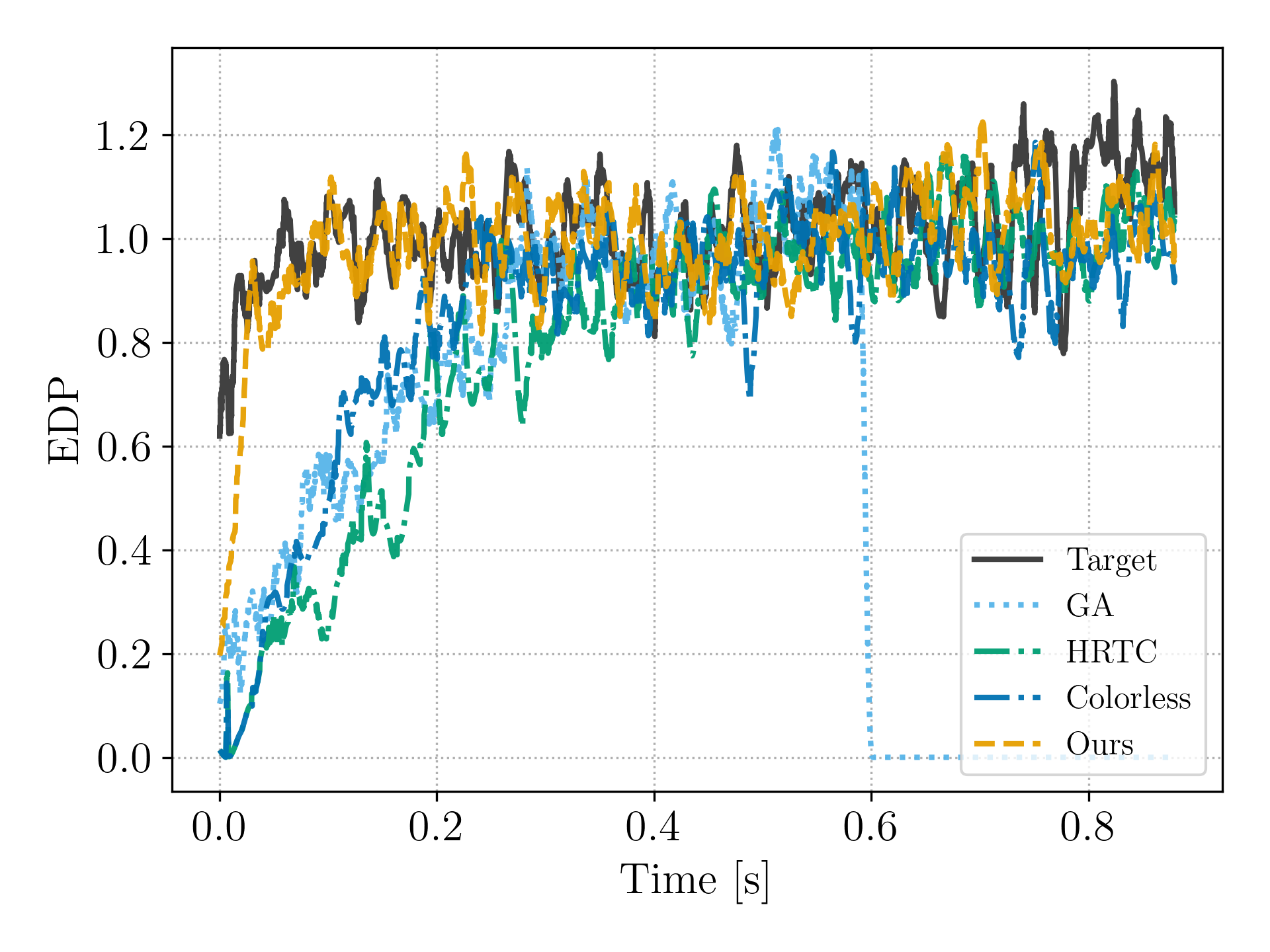}
    \vspace{-1em}
    \caption{Hallway (\texttt{h270}) EDPs.}
    \label{fig:edp_hallway}
\end{figure}

\begin{figure}[t]
    \centering
    \includegraphics[width=\linewidth]{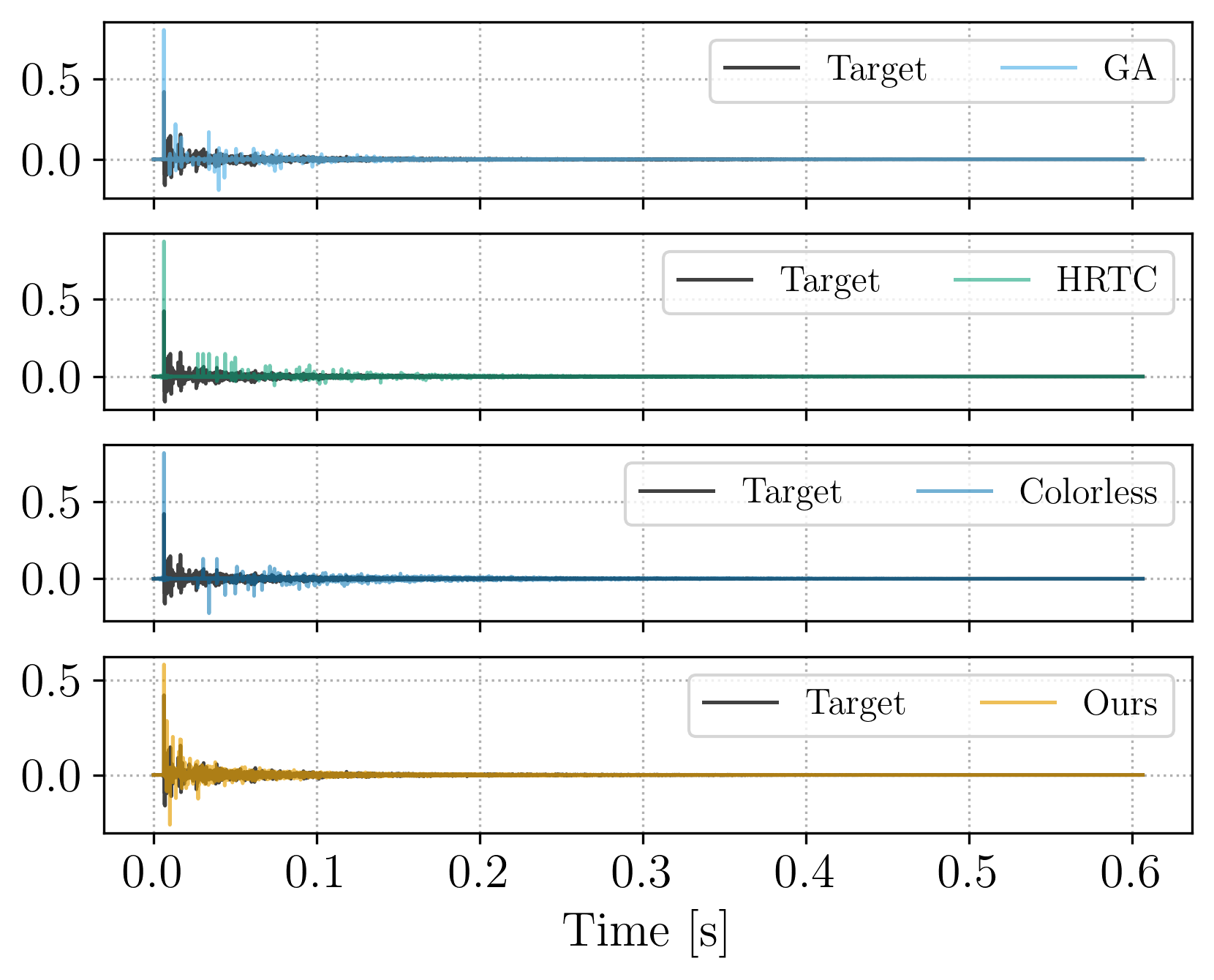}
    \vspace{-1em}
    \caption{Hallway (\texttt{h270}) IRs. {\color{black} The time axis is limited to the $T_{60}$ for visual clarity.}}
    \label{fig:rir_hallway}
\end{figure}

% ------------------------------------------

\pagebreak 

\subsection{Test Case: Hallway (h270)}
\label{ssec:test_case_hallway}

{\color{black}
Let us now consider the Hallway RIR (\texttt{h270}), which is characterized by nearly half the $T_{60}$ of the previous case.
Table~\ref{tab:loss_table} shows that the best results are obtained at iteration 796, where the loss is again lower by three orders of magnitude with respect to the starting point.

Figure~\ref{fig:edc_hallway} reports the EDCs of target, baselines, and proposed methods following the color conventions reported in the previous subsection. 
Once again, we evince the good matching between ours and the target decay, especially in the early and late portion of the curve.  
GA correctly matches the target EDC only in the first $100$~ms before rapidly decaying.
The HRTC and Colorless methods, instead, present a sharp energy drop after the direct path, which results in IRs characterized by an almost total absence of reflections for the first few ms, as shown in Figure~\ref{fig:rir_hallway}. HRTC, in turn, presents a slight overshoot that lasts for the first 300~ms, after which it closely matches the target EDC. 
%However, this happens when the energy has already reached $-30$~dB, thus after virtually all early reflections have occurred. 
The proposed method, instead, deviates from the target in its central part, approximately from 180 to 500~ms.
Arguably, however, these two kind of errors are not equivalent since the earlier portion of the RIR is known to be more relevant from a perceptual point of view~\cite{howard2013acoustics}

Figure~\ref{fig:edp_hallway} indicates that our approach is able to closely match the target EDP, further validating the effectiveness of the proposed Soft EDP loss function.
Indeed, the orange dashed curve follows the target curve until the $T_{60}$. We remind, in fact, that the training is performed only on such a span of time. Conversely, the EDP of the baseline methods show low values in the first 200~ms compared to the target one, and afterward, even if they take on comparable values, they do not follow the same trend.
This is confirmed by the RIRs shown in Figure~\ref{fig:rir_hallway}. Indeed, even in this test case, the FDN optimized by means of the proposed approach has an IR that resembles more of a realistic RIR, even though the amplitude of individual taps is not entirely matched.

Table~\ref{tab:metrics_hallway} reports the reverberation metrics. We can observe that the proposed optimization approach yields results comparable to the previous case, outperforming the baseline methods in five out of the six metrics.  
This time, due to the aforementioned mismatch in the central part of the EDCs in Figure~\ref{fig:edc_hallway}, HRTC and Colorless showcase a $\Delta T_{30}$ lower than that of the proposed method, with an error of $22.6$~ms and $54.6$~ms against the $85$~ms obtained with our approach.  
Oppositely, both HRTC and the proposed FDN render the $T_{60}$ equally well, with errors of $10.1$~ms and $9.2$~ms, respectively.
Our $\Delta C_{80}$ is one order of magnitude less than what can be achieved with the other methods. Likewise, our $\Delta D_{50}$ is one order of magnitude less than what can be achieved with GA and HRTC, and two orders of magnitude less than Colorless'. 
In addition, the center time error $\Delta t_{s}$ reaches $40.6$~\textmu s, while all baselines have errors one to three orders of magnitude larger, thus shifting the center of mass of the predicted IR energy more toward the reverberation tail.
}

% -------------------------------------------
\begin{figure}[t]
    \centering
    \includegraphics[width=\linewidth]{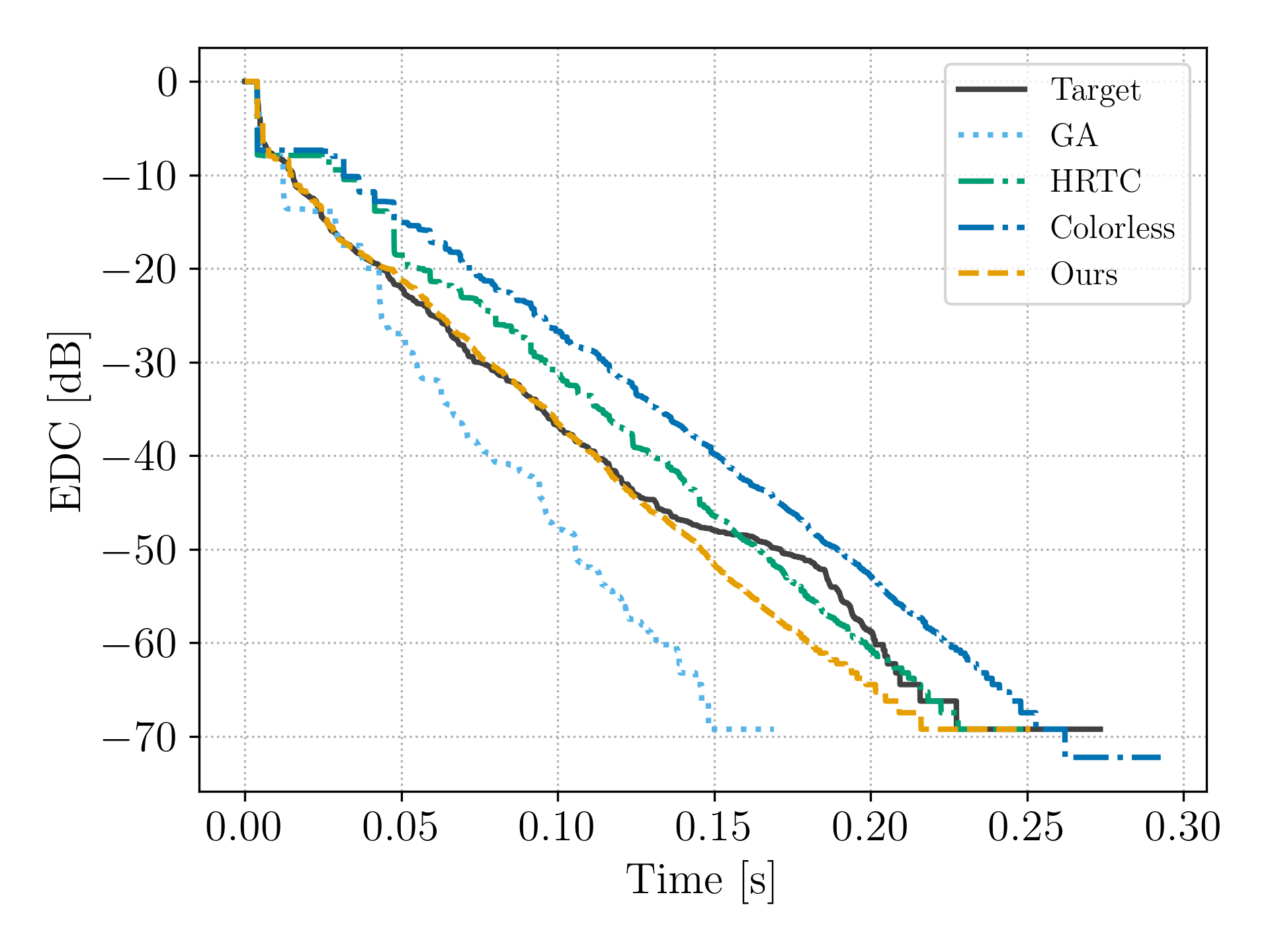}
    \vspace{-1em}
    \caption{Pizzeria (\texttt{h214}) EDCs.}
    \label{fig:edc_pizzeria}
\end{figure}
\begin{figure}[t]
    \centering
    \includegraphics[width=\linewidth]{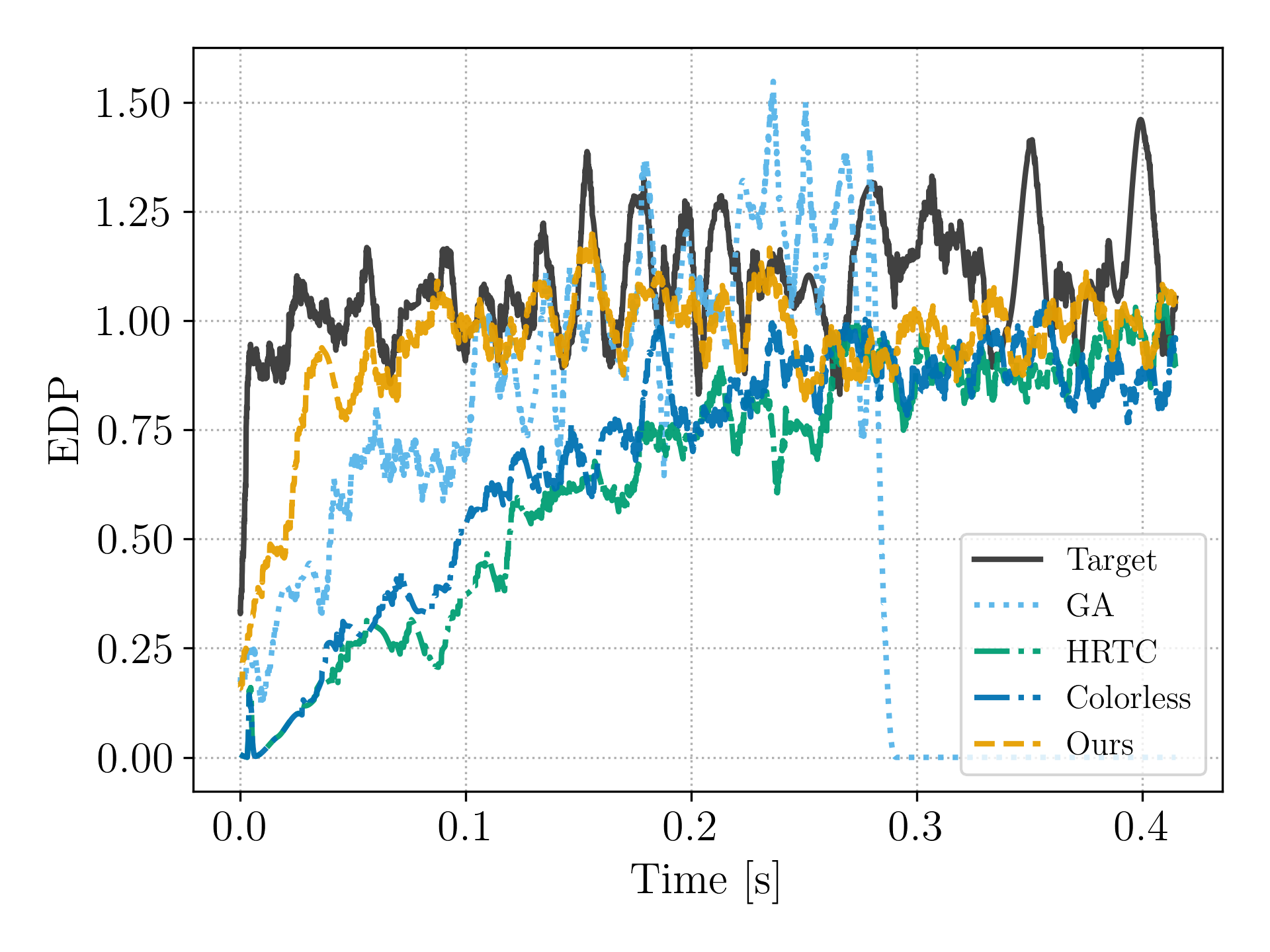}
    \vspace{-1em}
    \caption{Pizzeria (\texttt{h214}) EDPs.}
    \label{fig:edp_pizzeria}
\end{figure}

\begin{figure}[t]
    \centering
    \includegraphics[width=\linewidth]{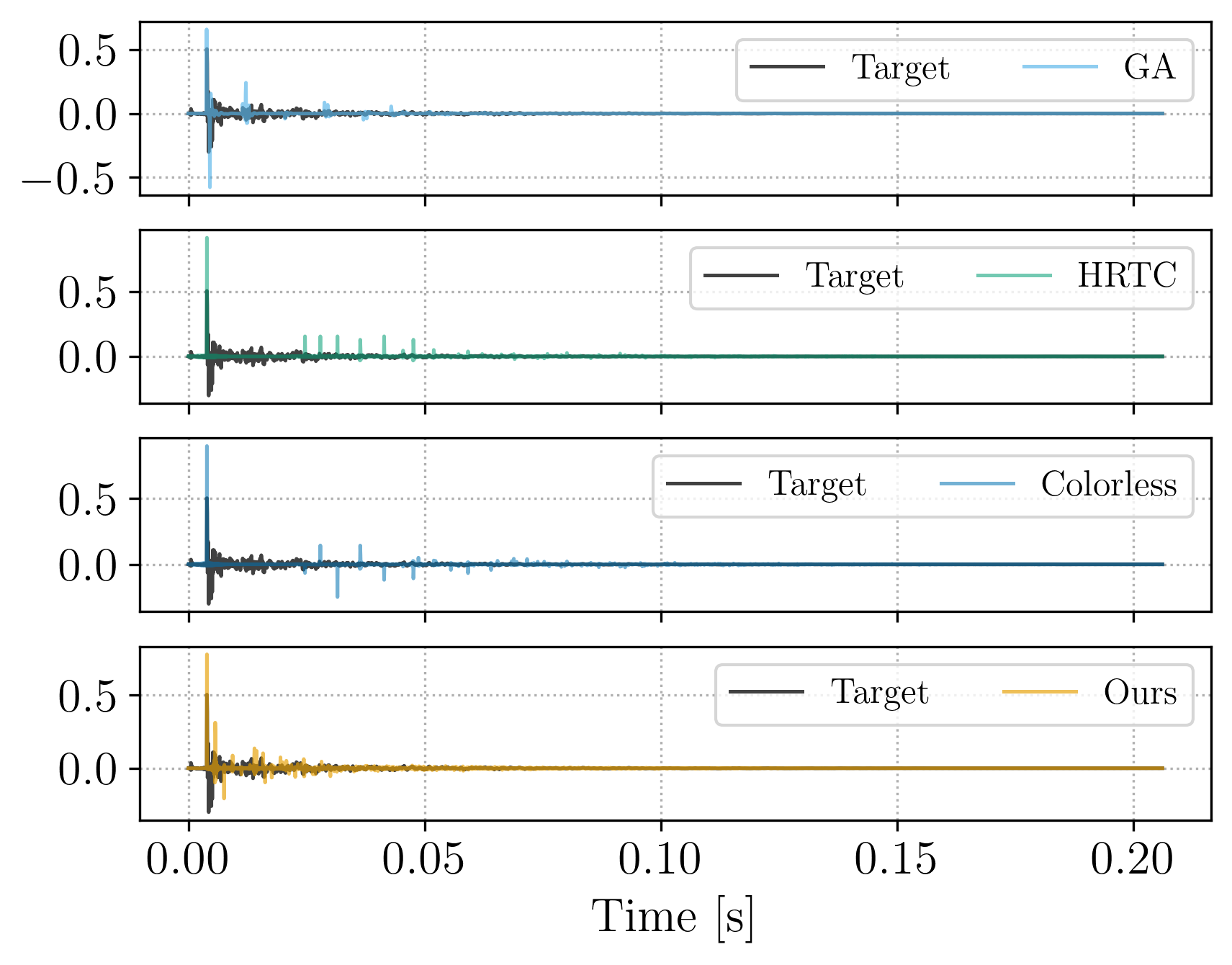}
    \vspace{-1em}
    \caption{Pizzeria (\texttt{h214}) IRs. {\color{black} The time axis is limited to the $T_{60}$ for visual clarity.}}
    \label{fig:rir_pizzeria}
\end{figure}
% -----------------------------------------

\subsection{Test Case: Pizzeria (h214)}
\label{ssec:test_case_pizzeria}
{\color{black}
Finally, let us focus on the shortest RIR of the three considered in the present study, i.e., \texttt{h214}, having a $T_{60}$ of just above 0.2~s.

Here, all the baseline methods appear to fail at modeling the target room (Figs.~\ref{fig:edc_pizzeria} and \ref{fig:edp_pizzeria}). In particular, the IRs shown in Figure~\ref{fig:rir_pizzeria} exhibit a few sparse taps, being thus far from resembling a real RIR. Since the backward-integrated energy abruptly decays with every peak, this results in a staircase-like behavior in the early portion of the EDC depicted in Figure~\ref{fig:edc_pizzeria}.

Facing the same difficulties, the proposed optimization method takes more gradient steps compared to the previous test case, converging at iteration 992 with a loss function two orders of magnitude lower than the starting value. 
In Figure~\ref{fig:edc_pizzeria}, the resulting EDC (dashed orange line) closely follows that of the target RIR (solid black line) until the two reach approximately $-45$~dB. Still, a direct comparison between EDCs in the range of $-45$ to $-70$~dB is not entirely reliable since the RIR has so little energy that background and sensor noise take on a much more relevant role when it comes to integrating the energy of the measured signal.

Overall, the proposed optimization method proves to perform well in fitting the RIR under scrutiny, with Table~\ref{tab:metrics_pizzeria} reporting an error of $4.7$~ms, $1.8$~ms, and $12.6$~ms when it comes to the three reverberation times. 
Furthermore,
$\Delta C_{80}$ is 0.41~dB, 
$\Delta D_{50}$ is $0.13$\%, 
and $\Delta t_s$ is $62.5$~\textmu s.
As in Section \ref{ssec:test_case_gym}, the only metric for which the proposed method shows a performance worse  than one of the baselines is the $T_{60}$. As a matter of fact, the HRTC is the only baseline to be once again characterized by a lower $\Delta T_{60}$, with a value of $5.2$~ms.
}

% -------------------------------------------

\subsection{Excluding the EDP Loss Term}
\label{ssec:ablation_edp_loss}
In developing our method, we noticed that using only the EDC loss function leads to ill-behaved IRs. Namely, we found that, while closely matching the desired EDC, the IR of an FDN trained with $\lambda=0$, i.e., using only $\mathcal{L}_\text{EDC}$, tends to exhibit an unrealistic echo distribution compared to the RIRs of real-life environments. In this section, we compare the results of our differentiable FDN trained without Soft EDP regularization with those presented in Section~\ref{ssec:test_case_hallway} obtained using the composite loss function in \eqref{eq:tot_loss}. For conciseness, we limit our analysis to the Hallway RIR (\texttt{h270}); results obtained with other RIRs are comparable to what is shown below.

\begin{figure}[t]
    \centering
    \subfloat[][]{\includegraphics[width=0.9\linewidth]{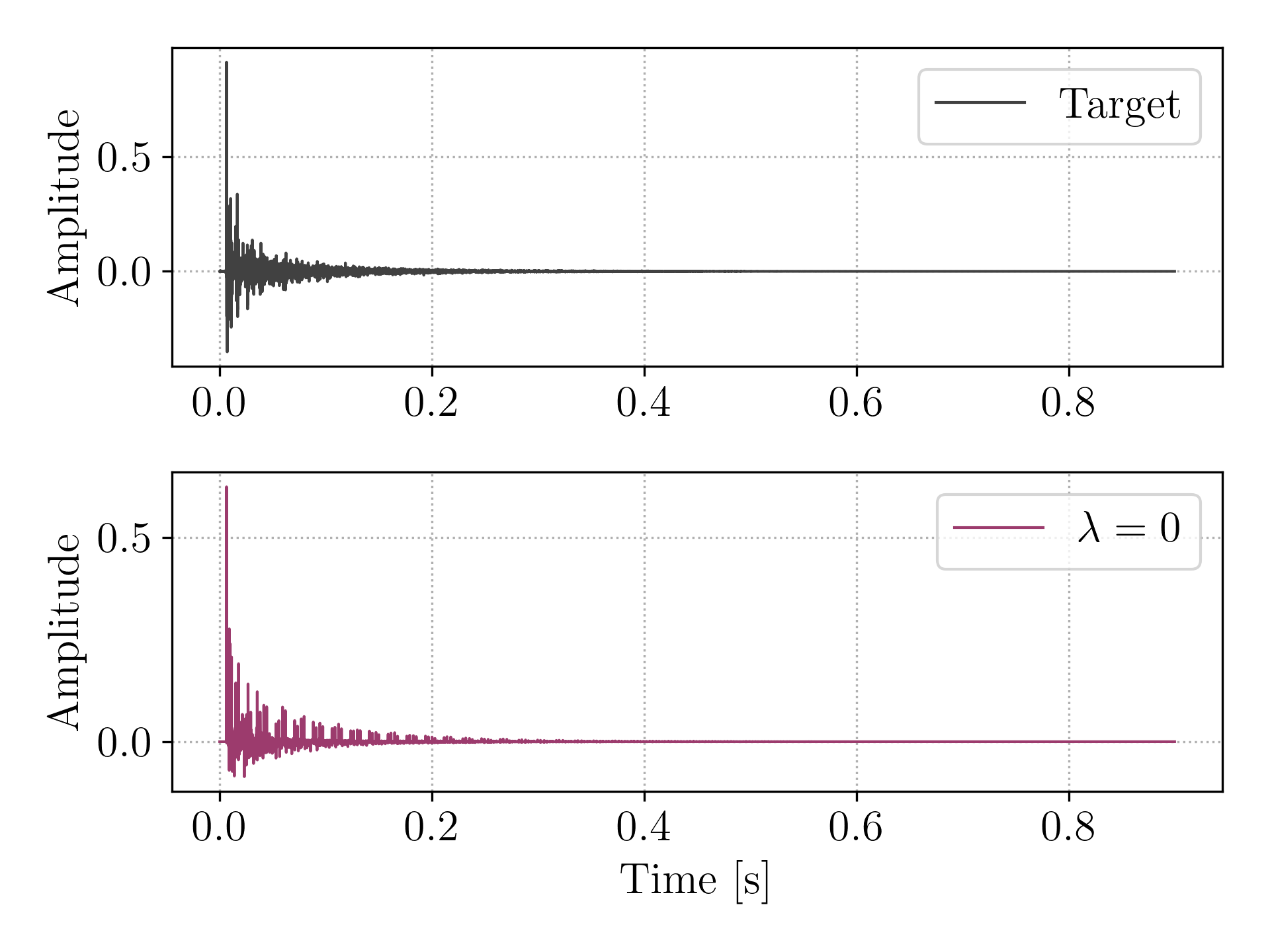}\label{fig:ablation_rir}}
    \hfill
    \vspace{-0.5em}
    \subfloat[][]{\includegraphics[width=\linewidth]{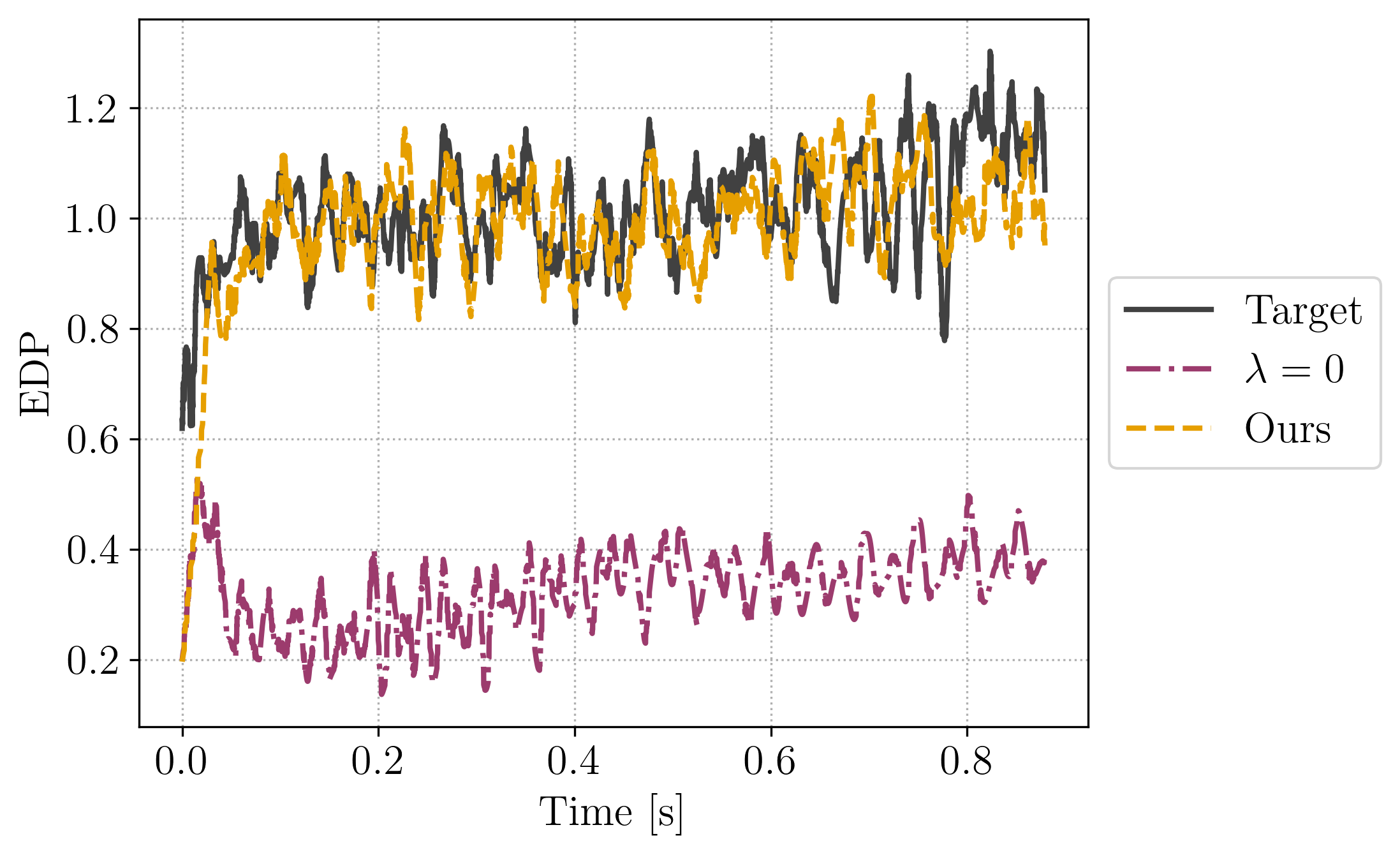}\label{fig:ablation_edp}}
    %\hfill
    \vspace{0.5em}
    \caption{(a) IR and (b) EDP obtained by training the FDN without EDP regularization term ($\lambda=0$).}
    \label{fig:ablation}
\end{figure}

{\color{black}
When excluding the EDP loss term from~\eqref{eq:tot_loss}, the NMSE between true and predicted EDCs~\eqref{eq:edc_loss} is comparable with that of the proposed method, totaling $2.8\times10^{-3}$ ($\lambda=0$) and $3.4\times10^{-3}$ ($\lambda=0.1$), respectively.
Yet, the MSE between true and predicted EDPs~\eqref{eq:edp_loss} is $0.342$, i.e., two orders of magnitude higher than the $6.8\times10^{-3}$ reported in Table~\ref{tab:loss_table} for the proposed method. This can be observed in Figure~\ref{fig:ablation_edp}, showing that the echo density when $\lambda=0$ (dash-dotted purple line) is far from the desired profile (solid black line). The target RIR and the proposed method (dashed orange line) produce an EDP with values consistently around one, indicating dense reverberation. On the contrary, the FDN trained without Soft EDP regularization yields an EDP with values below $0.5$, signaling an uneven echo density due to the presence of a few prominent reflections~\cite{huang2007aspects}.
This observation is confirmed by the IR shown in Figure~\ref{fig:ablation_rir} that exhibits an exponentially decaying cluster of four taps periodically peaking above a denser reverberation tail with negative polarity.

Excluding the EDP loss term means solving a minimization problem with no constraints discouraging the model to use just a small number of delay lines to capture the overall EDC behavior.
Since FDNs extend a parallel comb-filter structure, we believe that the behavior observed in Figure~\ref{fig:ablation_rir} is related to the well-understood problem occurring when only some delay lines are strongly excited and recirculated, which, in turn, aggravates the comb-like behavior of the delay network, ultimately resulting in an unpleasing metallic sound quality~\cite{schroeder1961colorless, dalsanto2023colorless}.
}

\subsection{Soft EDP Approximation}
\label{ssec:ablation_kappa}
\begin{figure}[t]
    \centering
    \subfloat[][Pizzeria (\texttt{h214})]{\includegraphics[width=\linewidth]{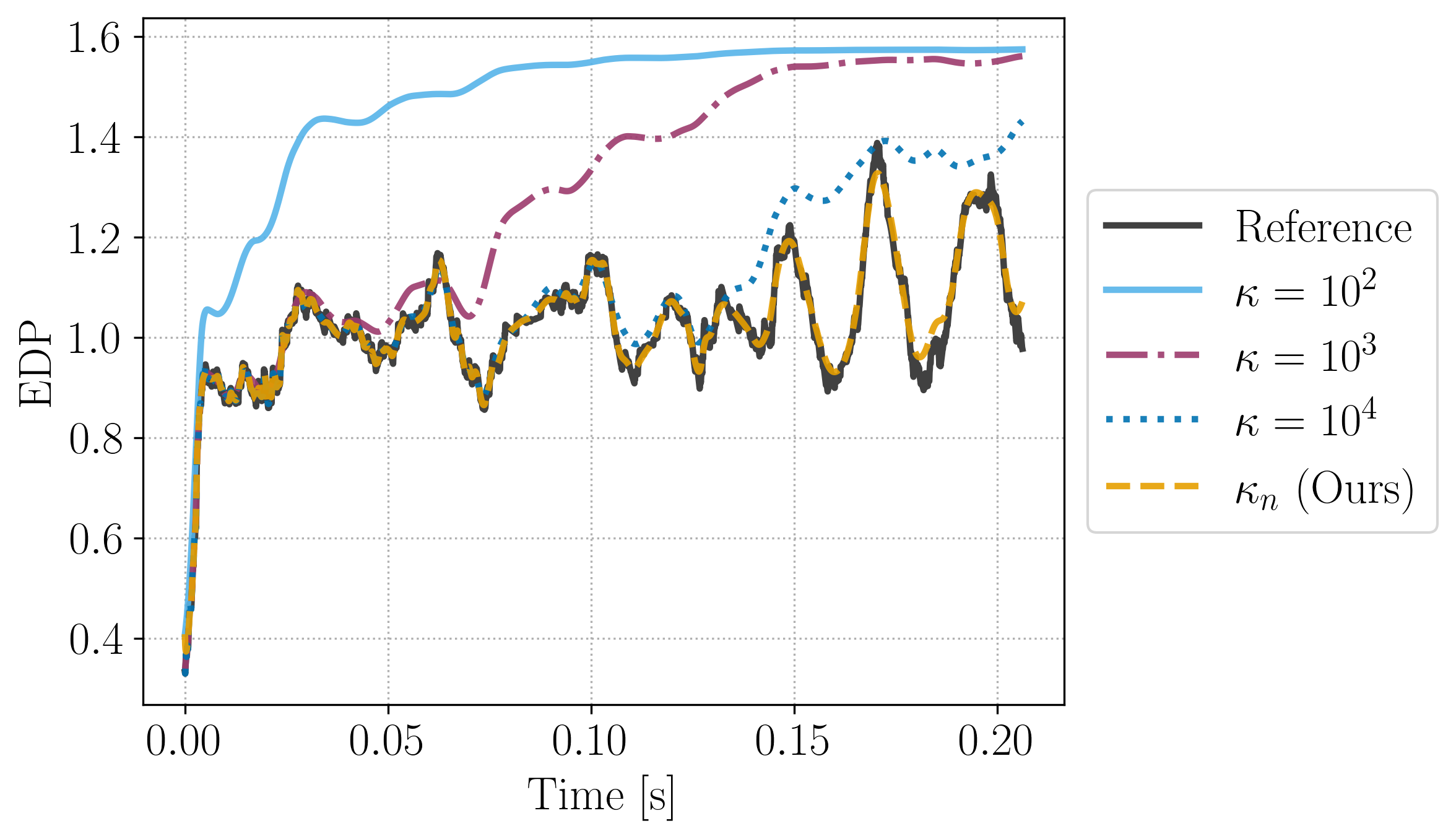}\label{fig:soft_edp_pizzeria}}\hfill
    \subfloat[][Hallway (\texttt{h270})]{\includegraphics[width=\linewidth]{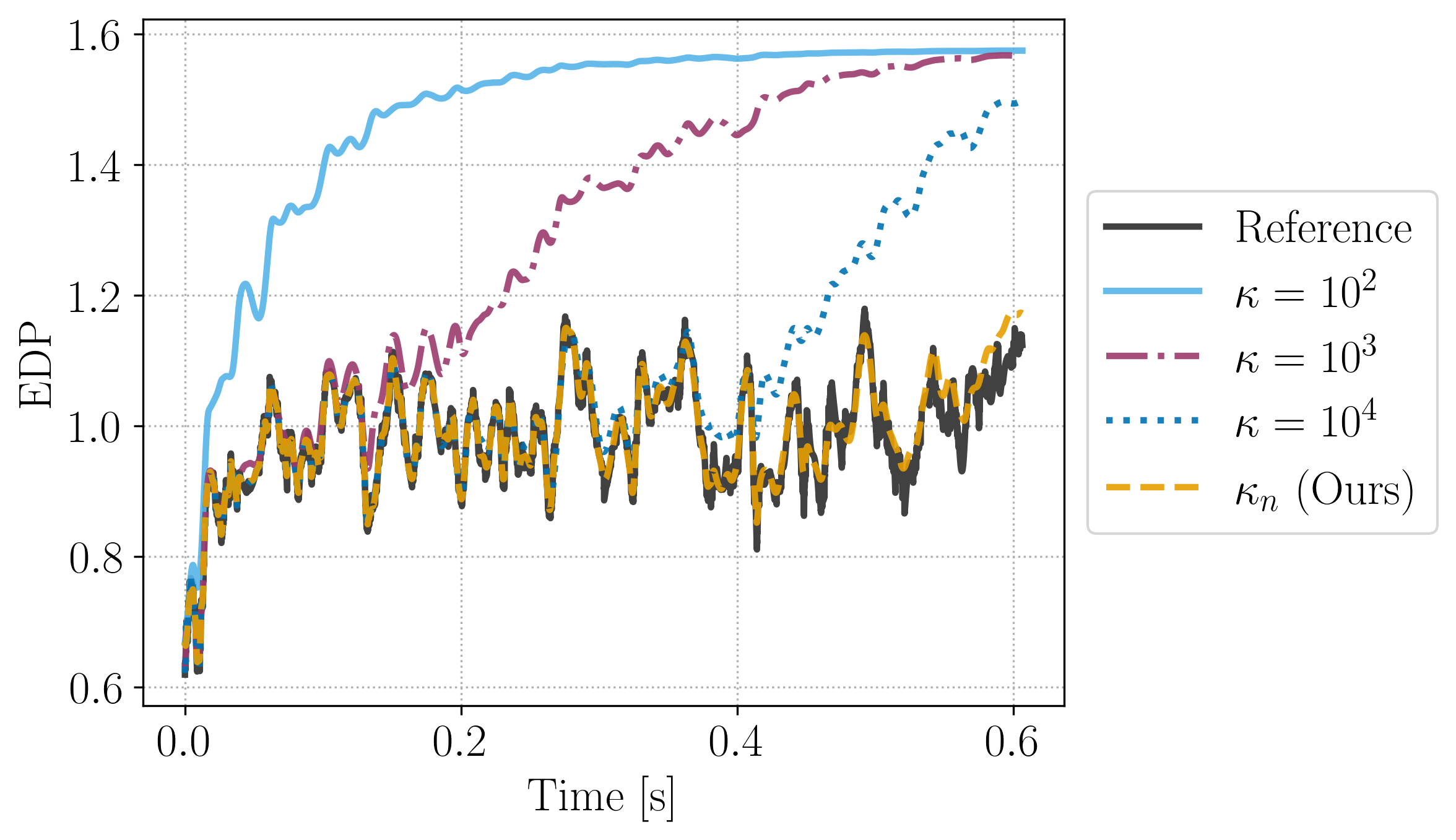}\label{fig:soft_edp_hallway}}\hfill
    \subfloat[][Gym (\texttt{h052})]{\includegraphics[width=\linewidth]{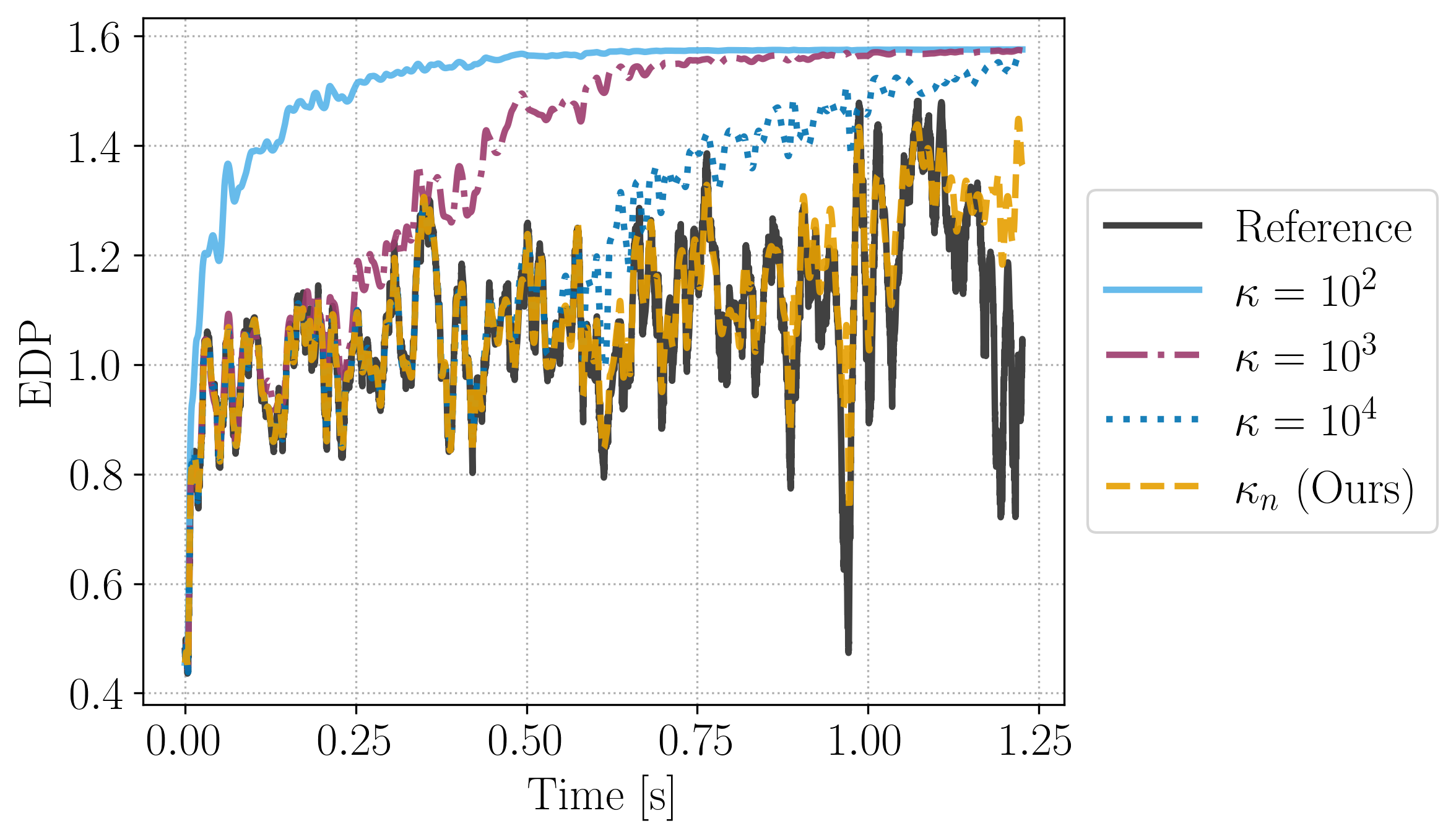}\label{fig:soft_edp_gym}}\hfill
    \vspace{1em}
    \caption{Non-differentiable EDP (Reference) compared with the proposed Soft EDP function for different values of the Sigmoid scaling parameter.}
    \label{fig:soft_edp_approx}
\end{figure}

Finally, we discuss the approximation capabilities of the proposed Soft EDP function introduced in Section~\ref{sssec:edp_loss}. Figure~\ref{fig:soft_edp_approx} shows the non-differentiable EDP (solid black line) defined in \eqref{eq:edp} against several Soft EDP approximations of the three RIRs considered in the present study. 
We test various scaling parameters $\kappa$, namely, $10^2$, $10^3$, and $10^4$, along with the proposed time-varying $\kappa_n$ linearly increasing from $10^2$ ($n=0$) to $10^5$ ($n=L_{T_{60}} - 1$). We depict the profiles only for time indices below the $T_{60}$, as this range is the one considered when training the FDNs.

In Figure~\ref{fig:soft_edp_approx}, we may notice that $\kappa=10^2$ yields a poor approximation of the reference EDP beyond the very first few ms.
We also observe that $\kappa=10^3$ and $\kappa=10^4$ provide relative improvements. However, after some time, the approximation starts to degrade in a similar fashion as for $\kappa=10^2$. Conversely, the proposed Soft EDP with time-varying scaling (dashed orange line) is able to closely match the non-differentiable reference profile all the way up to the $T_{60}$ in Figure~\ref{fig:soft_edp_pizzeria}, while gracefully combating gradient vanishing (see Section~\ref{sssec:edp_loss}). Notably, however, Figure~\ref{fig:soft_edp_gym} shows that the approximation in the very last portion of the longest RIR considered, i.e., Gym (\texttt{h052}), significantly differs from the reference profile. To a lesser extent, this is also noticeable in Figure~\ref{fig:soft_edp_hallway}. Nevertheless, it is worth mentioning that, around the estimated $T_{60}$, the energy of the RIR is already almost entirely vanished, and the EDP itself suffers from statistical uncertainty due to sensor and 16-bit quantization noise.

% -------------------------------------------
 
% -------------------------------------------

\begin{table}[t]
\caption{Metrics for the Gym RIR.}\label{tab:metrics_gym}%
\centering
\setlength{\tabcolsep}{0pt}
\begin{tabular*}{\linewidth}{@{\extracolsep{\fill}}lcccccc@{}}
\toprule
& $T_{20}$ & $\Delta T_{20}$ & $T_{30}$ & $\Delta T_{30}$ & $T_{60}$ & $\Delta T_{60}$\\
\midrule
Target & $0.8616$ & --- & $0.9161$ & --- & $1.2257$ & --- \\
GA & $0.6334$ & $0.2282$ & $0.6969$ & $0.2193$ & $0.8117$ & $0.4139$ \\
HRTC & $1.2566$ & $0.3951$ & $1.2597$ & $0.3436$ & $1.2194$ & $\underline{0.0063}$ \\
Colorless & $1.2589$ & $0.3973$ & $1.2504$ & $0.3342$ & $1.1900$ & $0.0357$ \\
Ours & $0.8451$ & $\underline{0.0165}$ & $0.9714$ & $\underline{0.0552}$ & $1.1355$ & $0.0902$ \\
\end{tabular*}
\setlength{\tabcolsep}{0pt}
\begin{tabular*}{\linewidth}{@{\extracolsep{\fill}}lcccccc@{}}
\toprule
& $C_{80}$ &  $\Delta C_{80}$&  $D_{50}$ &  $\Delta D_{50}$ & $t_s$ & $\Delta t_s$ \\
\midrule
Target & $12.106$ & --- & $89.009$ & --- & $22.899$ & --- \\
GA & $11.985$ & $0.1208$ & $91.194$ & $2.1848$ & $19.343$ & $3.5560$ \\
HRTC & $7.8183$ & $4.2877$ & $79.402$ & $9.6070$ & $38.346$ & $15.447$ \\
Colorless & $2.7802$ & $9.3258$ & $56.329$ & $32.679$ & $77.035$ & $54.135$ \\
Ours & $12.126$ & $\underline{0.0200}$ & $88.912$ & $\underline{0.0974}$ & $23.079$ & $\underline{0.1805}$ \\
 \bottomrule
\end{tabular*}
\end{table}
%
%\vspace{2em}
%
\begin{table}[t]
\caption{Metrics for the Hallway RIR.}\label{tab:metrics_hallway}%
\centering
\setlength{\tabcolsep}{0pt}
\begin{tabular*}{\linewidth}{@{\extracolsep{\fill}}lcccccc@{}}
\toprule
& $T_{20}$ & $\Delta T_{20}$ & $T_{30}$ & $\Delta T_{30}$ & $T_{60}$ & $\Delta T_{60}$\\
\midrule
Target & $0.5289$ & --- & $0.6010$ & --- & $0.6067$ & --- \\
GA & $0.4200$ & $0.1089$ & $0.4233$ & $0.1777$ & $0.4329$ & $0.1738$ \\
HRTC & $0.6362$ & $0.1072$ & $0.6236$ & $\underline{0.0226}$ & $0.6168$ & $0.0101$ \\
Colorless & $0.7007$ & $0.1717$ & $0.6556$ & $0.0546$ & $0.6335$ & $0.0268$ \\
Ours & $0.4749$ & $\underline{0.0540}$ & $0.5160$ & $0.0850$ & $0.5975$ & $\underline{0.0092}$ \\
\end{tabular*}
\setlength{\tabcolsep}{0pt}
\begin{tabular*}{\linewidth}{@{\extracolsep{\fill}}lcccccc@{}}
\toprule
& $C_{80}$ &  $\Delta C_{80}$&  $D_{50}$ &  $\Delta D_{50}$ & $t_s$ & $\Delta t_s$ \\
\midrule
Target & $14.079$ & --- & $90.691$ & --- & $20.116$ & --- \\
GA & $15.502$ & $1.4223$ & $92.106$ & $1.4147$ & $18.365$ & $1.7514$ \\
HRTC & $11.412$ & $2.6678$ & $88.246$ & $2.4453$ & $20.793$ & $0.6767$ \\
Colorless & $8.327$ & $5.7520$ & $77.850$ & $12.841$ & $31.269$ & $11.152$ \\
Ours & $13.759$ & $\underline{0.3204}$ & $90.856$ & $\underline{0.1648}$ & $20.075$ & $\underline{0.0406}$ \\
 \bottomrule
\end{tabular*}
\end{table}
%
%\vspace{2em}
%
\begin{table}[t]
\caption{Metrics for the Pizzeria RIR.}\label{tab:metrics_pizzeria}%
\centering
\setlength{\tabcolsep}{0pt}
\begin{tabular*}{\linewidth}{@{\extracolsep{\fill}}lcccccc@{}}
\toprule
& $T_{20}$ & $\Delta T_{20}$ & $T_{30}$ & $\Delta T_{30}$ & $T_{60}$ & $\Delta T_{60}$\\
\midrule
Target & $0.1643$ & --- & $0.1794$ & --- & $0.2062$ & --- \\
GA & $0.1172$ & $0.0471$ & $0.1235$ & $0.0559$ & $0.1417$ & $0.0645$ \\
HRTC & $0.2286$ & $0.0643$ & $0.2185$ & $0.0391$ & $0.2114$ & $\underline{0.0052}$ \\
Colorless & $0.2668$ & $0.1026$ & $0.2557$ & $0.0764$ & $0.2367$ & $0.0306$ \\
Ours & $0.1689$ & $\underline{0.0047}$ & $0.1811$ & $\underline{0.0018}$ & $0.1936$ & $0.0126$ \\
\end{tabular*}
\setlength{\tabcolsep}{0pt}
\begin{tabular*}{\linewidth}{@{\extracolsep{\fill}}lcccccc@{}}
\toprule
& $C_{80}$ &  $\Delta C_{80}$&  $D_{50}$ &  $\Delta D_{50}$ & $t_s$ & $\Delta t_s$ \\
\midrule
Target & $30.988$ & --- & $99.382$ & --- & $7.1698$ & --- \\
GA & $40.683$ & $9.6950$ & $99.799$ & $0.4170$ & $6.4524$ & $0.7174$ \\
HRTC & $24.860$ & $6.1277$ & $98.611$ & $0.7708$ & $9.4454$ & $2.2756$ \\
Colorless & $21.849$ & $9.1389$ & $96.895$ & $2.4862$ & $10.749$ & $3.5798$ \\
Ours & $30.576$ & $\underline{0.4123}$ & $99.250$ & $\underline{0.1322}$ & $7.2323$ & $\underline{0.0625}$ \\
 \bottomrule
\end{tabular*}
\end{table}

% -------------------------------------------

\subsection{Limitations}
\begin{figure*}
    \centering
    \subfloat[][Target]{\includegraphics[width=0.25\linewidth]{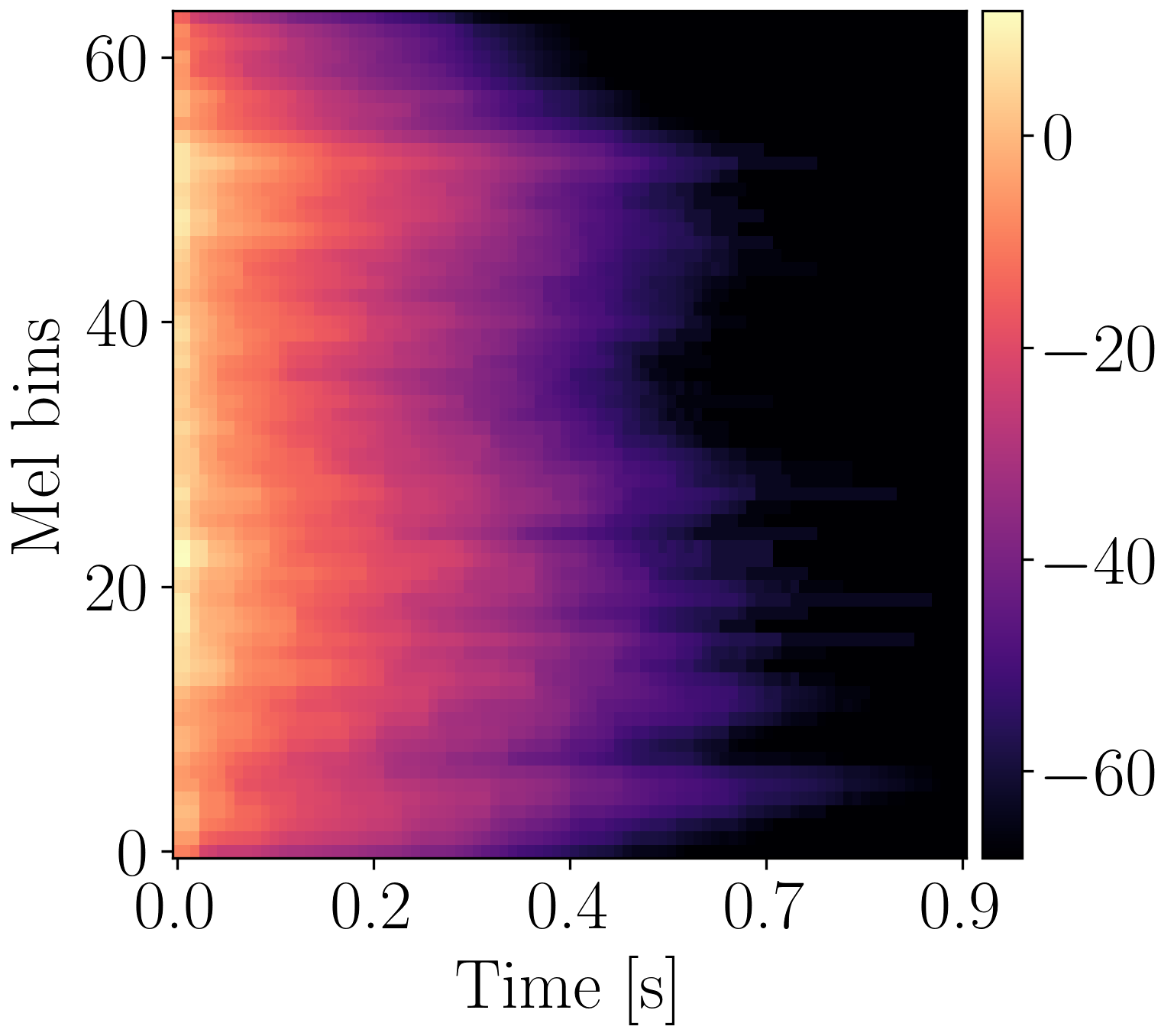}}
    \hspace{1em}
    \subfloat[][HRTC]{\includegraphics[width=0.25\linewidth]{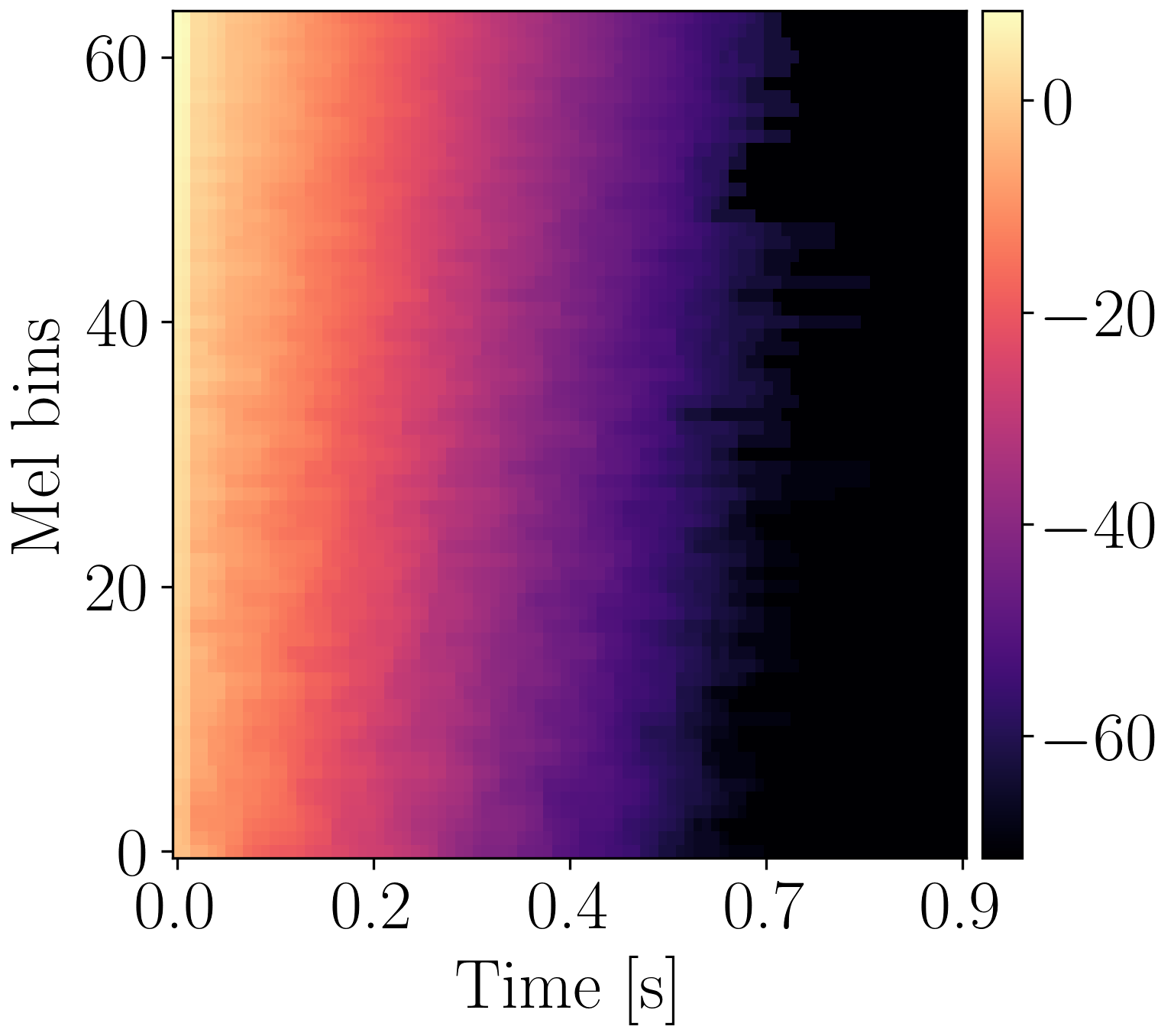}}
    \hspace{1em}
    \subfloat[][GA]{\includegraphics[width=0.25\linewidth]{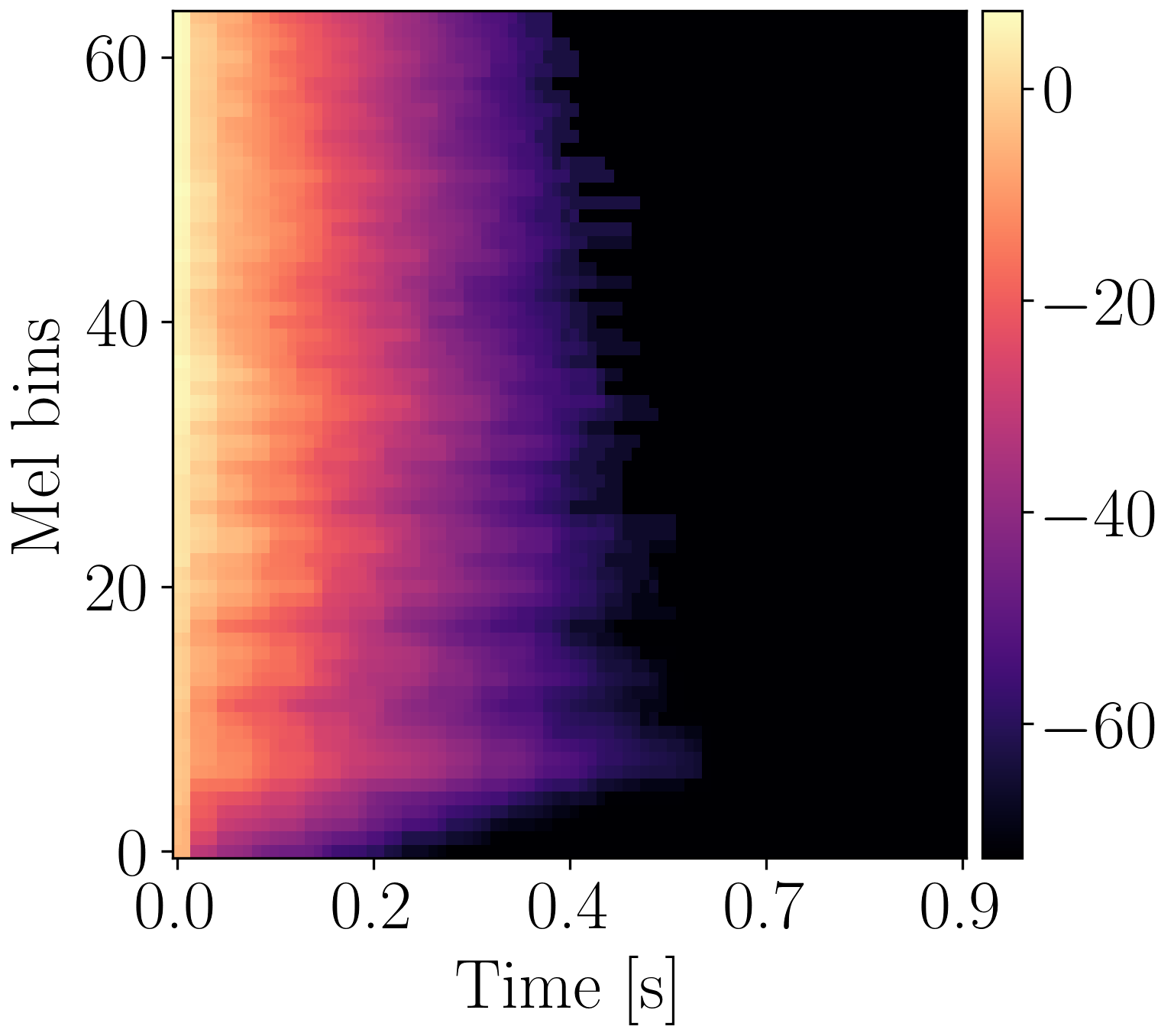}}\\
    \subfloat[][Colorless]{\includegraphics[width=0.25\linewidth]{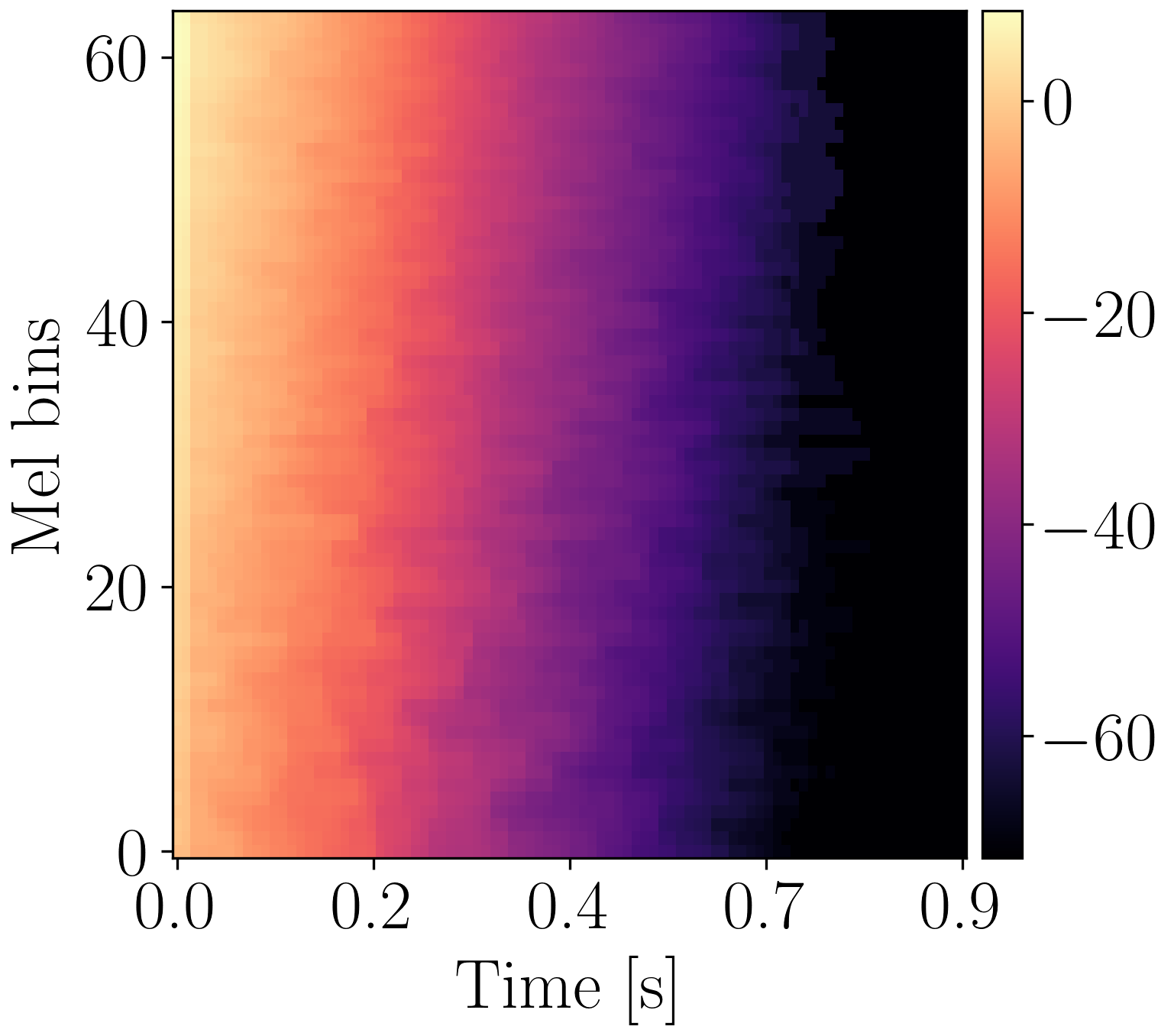}}
    \hspace{1em}
    \subfloat[][Ours]{\includegraphics[width=0.25\linewidth]{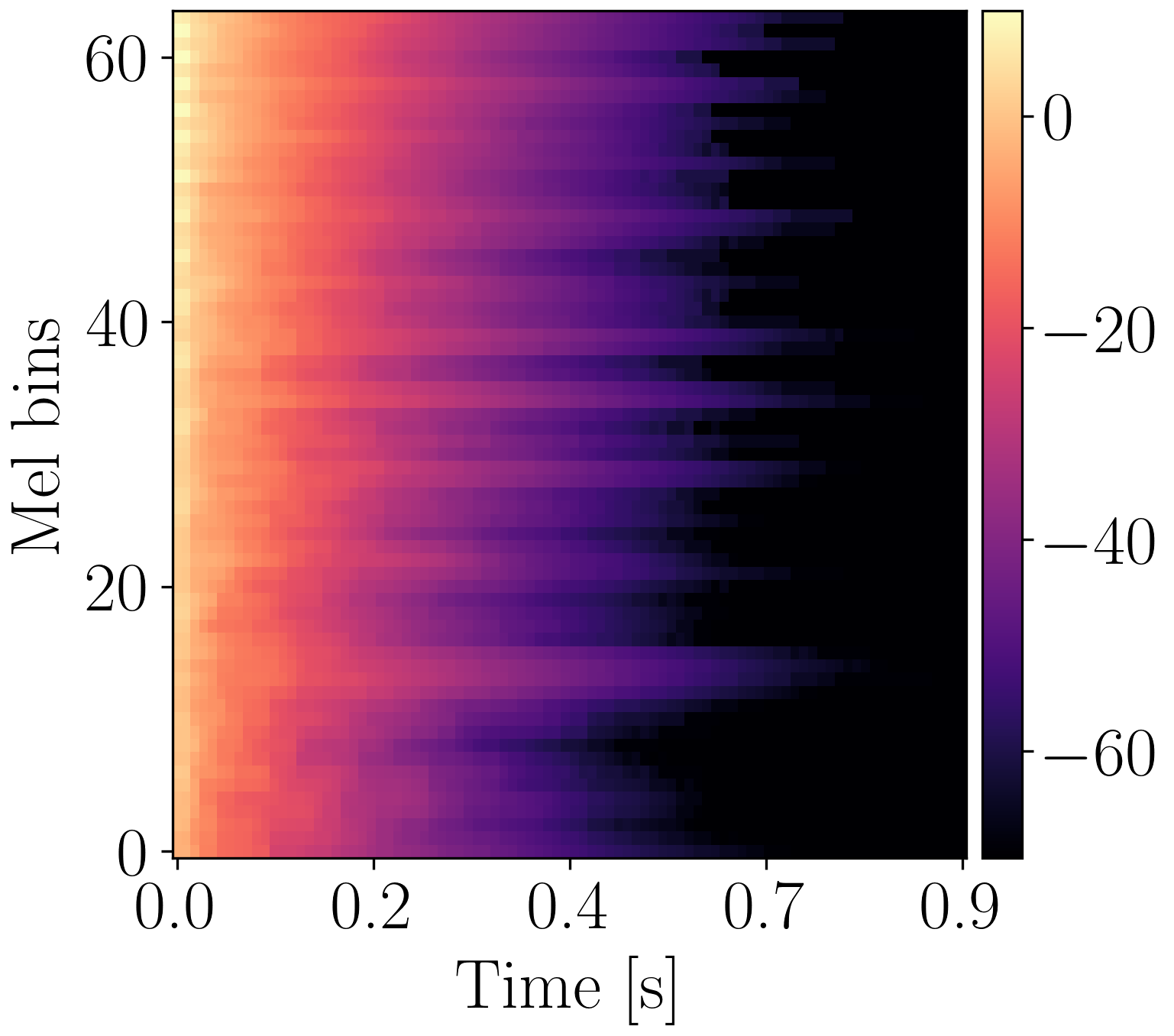}}
    \caption{{\color{black} Mel-frequency Energy Decay Reliefs (EDRs) of the Hallway test case (\texttt{h270}). Proposed in~\cite{jot1992analysis}, the EDR extends Schroeder's EDC to multiple frequency bands. Here, the 512-bin frequency axis is warped onto a 64-bin mel scale.}}
    \label{fig:edr-limitations}
\end{figure*}

{\color{black}
In this work, we focused on two important perceptual characteristics of room impulse responses: integrated energy decay and echo density. While rendering these time-domain features is key for any artificial reverberation algorithm that aims to be realistic, by themselves, Schroeder’s EDC and Abel and Huang's EDP are not enough to comprehensively model the perceptual qualities of reverberation. 
In fact, it is well known that frequency- and time-frequency features play a crucial role in room acoustic simulation~\cite{fifty_years, mit2016}. 
%\cite{jot_fdn, jot1992analysis, mit2016}.

However, Figures \ref{fig:gym_freqz}, \ref{fig:hallway_freqz}, \ref{fig:pizzeria_freqz} show that none of the FDNs considered in the present study manages to capture the magnitude frequency response of the target RIRs. Likewise, Figure~\ref{fig:edr-limitations} reveals a significant discrepancy between the target Energy Decay Relief (EDR) \cite{jot1992analysis} and those of the FDN models.

Such a conspicuous mismatch entails that the output signals of the FDNs sound different not only from one another but also from the corresponding input signal convolved with the target RIR.
In turn, this undermines the reliability of any perceptual test assigning similarity scores to each method with respect to the target, as subjective judgments would be significantly influenced by differing spectro-temporal coloration and decay. In this respect, pilot experiments proved inconclusive, highlighting the need for further investigation into time-frequency modeling.

After all, matching the spectro-temporal characteristics of the target RIRs is not among the objectives of the parameter selection/tuning algorithms considered in the present study. %Still, despite considering cepstral features, GA is far from resembling the reference spectro-temporal behavior.
%the magnitude of the resulting transfer function is far from resembling the target spectrum.
Moreover, in this work, we mainly focused on time-invariant frequency-independent FDN prototypes.
This holds true for proposed and baseline methods but GA, whose fitness function consists of a $L^1$-loss between MFCCs. Still, despite considering cepstral features, the result obtained with GA is far from resembling the reference spectro-temporal behavior. This evidences that it is not straightforward to accurately capture both time- and frequency-domain characteristics through an optimization process, even when including dedicated absorption and tone correction filters in the FDN prototype (cf.~Figure~\ref{fig:baseline_fdn}). 

Although said filters can be implemented in a differentiable fashion, the poor performance of GA points out that extending the cost function \eqref{eq:tot_loss} to the frequency-dependent case may not be sufficient, ultimately suggesting that a more thorough and comprehensive study is necessary to accomplish the goal. We leave such an investigation for future work.
%Although absorption and tone correction filters can be implemented in a differentiable fashion, and it is possible to extend the cost function to the frequency-dependent case by adding a regression loss term in \eqref{eq:tot_loss}, the poor performance of GA suggests that a more thorough and comprehensive study is necessary. We leave such an investigation for future work.

%placing a filter having the desired magnitude response at the output of the FDN in order to correct the system's power spectral density.
%
%The design of tone correction filters, however, besides being outside the scope of the present work, cannot resolve potential time-dependent mismatches in the EDR. For this purpose, in fact, a modification to the FDN prototype itself would be needed \cite{jot1992analysis}. Moreover, in the case of strongly inhomogeneous decay across frequency bands, cascading an equalizer may risk affecting Schroeder’s integrated EDC as well.

%is a practical solution for those IRs where all frequencies decay at the same rate, to the best of our knowledge, there is no easy way to decouple spectral coloration from sound energy decay in the case of inhomogeneous decay across frequency bands.

%Following classic FDN design principles~\cite{jot_fdn}, one might think of augmenting the FDN prototype to include absorption and tone correction filters as shown in Figure~\ref{fig:baseline_fdn}.
%
%In principle, . A thorough analysis of this approach is left for future work.

\begin{figure}
    \centering
    \includegraphics[width=\linewidth]{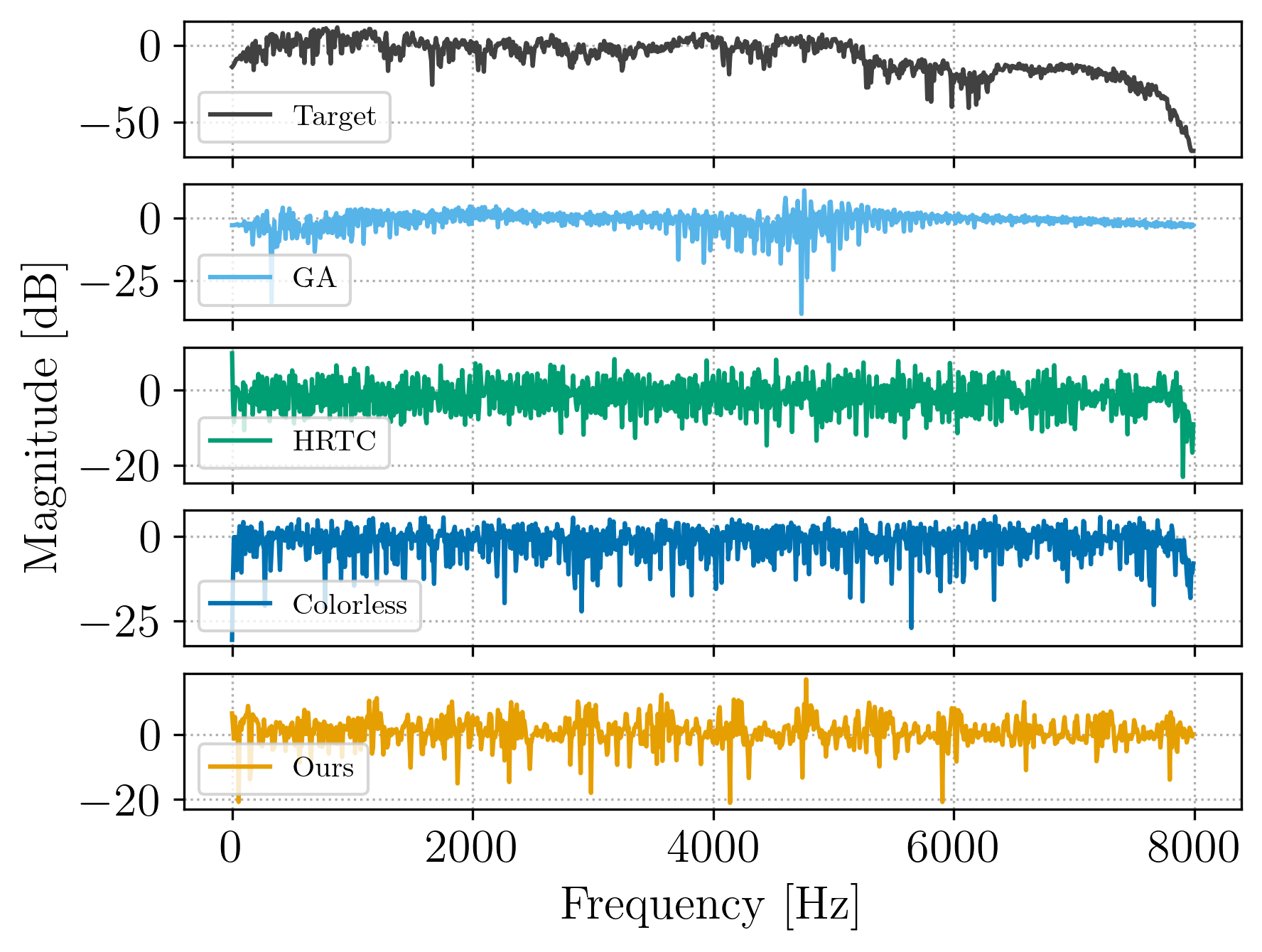}
    \vspace{-1em}
    \caption{Magnitude of the frequency response of the Gym RIR (\texttt{h052}) and corresponding FDN transfer functions.}
    \label{fig:gym_freqz}
    \vspace{1em}
    \includegraphics[width=\linewidth]{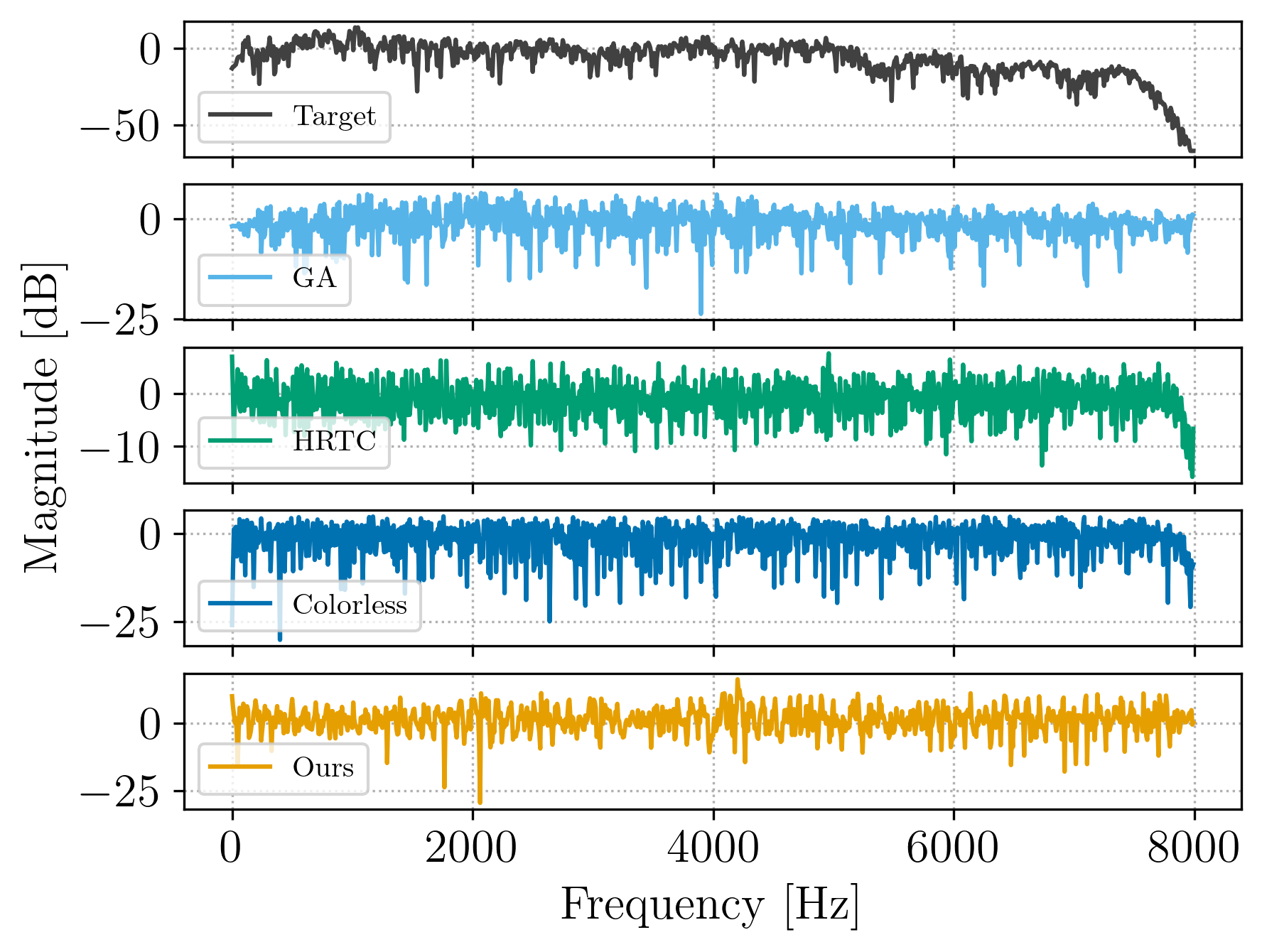}
    \vspace{-1em}
    \caption{Magnitude of the frequency response of the Hallway RIR (\texttt{h270}) and corresponding FDN transfer functions.}
    \label{fig:hallway_freqz}
    \vspace{1em}
    \includegraphics[width=\linewidth]{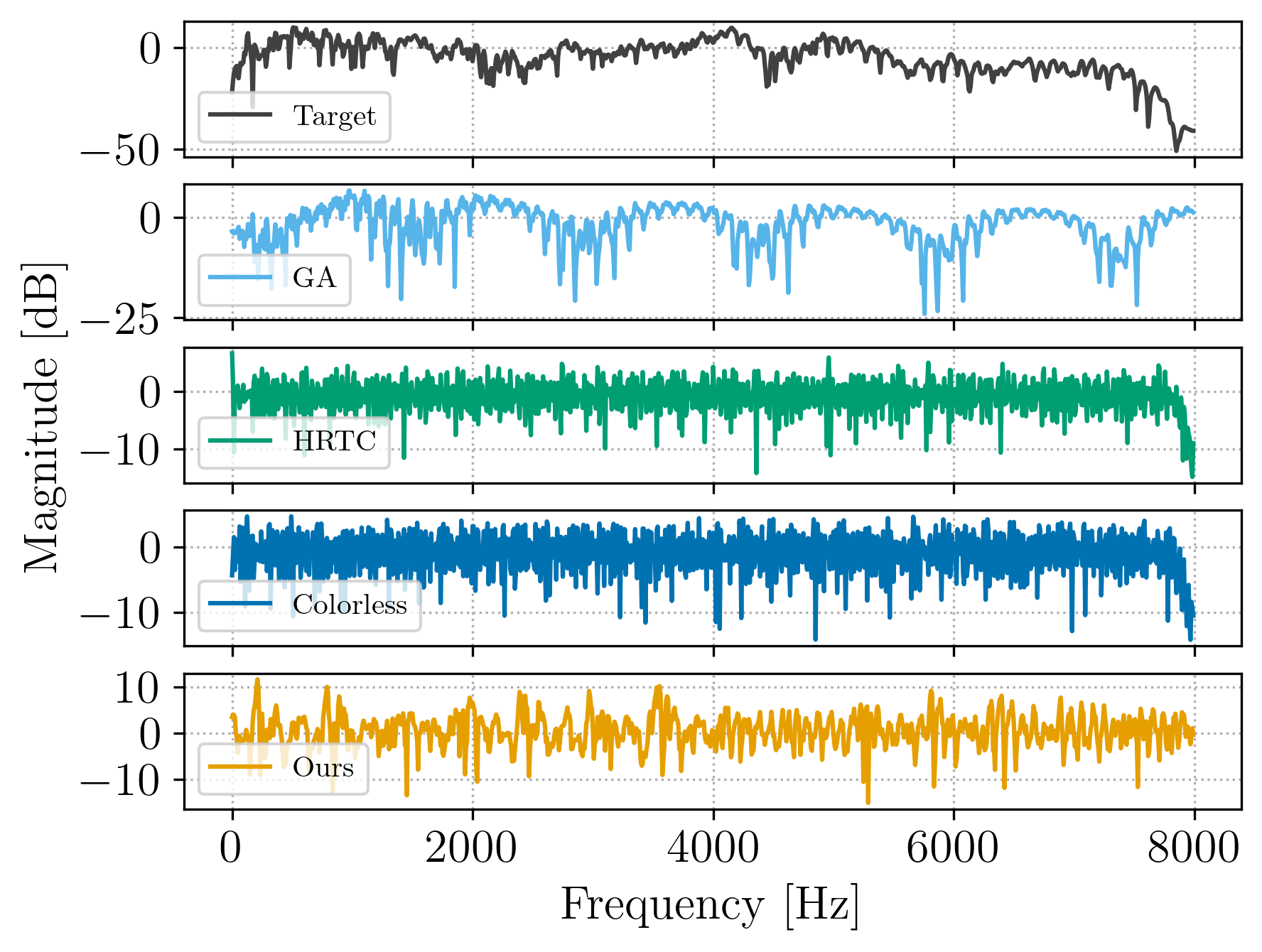 }
    \vspace{-1em}
    \caption{Magnitude of the frequency response of the Pizzeria RIR (\texttt{h214}) and corresponding FDN transfer functions.}
    \label{fig:pizzeria_freqz}
\end{figure}
}

\section{Conclusions}
\label{ref:conclusions}
In this work, we proposed a method for optimizing every parameter of a time-invariant frequency-independent Feedback Delay Network (FDN) so as to match the reverberation of a given room through perceptually meaningful metrics. 
%
%In the present work,
{\color{black} The main contributions are the following:}
\begin{itemize}
    \item[i.] We introduced a differentiable FDN with learnable delay lines,
    \item[ii.] We developed a novel optimization framework for \textit{all} FDN parameters based on automatic differentiation,
    \item[iii.] We applied gradient-based optimization with the objective of matching selected acoustic features of measured RIRs,
    \item[iv.] We presented an innovative use of established perceptually-motivated acoustic measures as loss terms, 
    \item[v.] We proposed a differentiable approximation of the well-known normalized Echo Density Profile named Soft EDP.
\end{itemize}

\noindent In particular, we presented a new differentiable FDN that is characterized by learnable delay lines realized exploiting operations in the frequency domain.
Thus, we jointly trained all FDN parameters via backpropagation taking into account a composite loss consisting of two terms: the normalized mean square error between target and predicted backward-integrated EDCs and the mean square error between target and predicted Soft EDPs. We evaluated the proposed method on three real-world RIRs taken from a publicly available dataset, and we demonstrated that the Soft EDP term is essential for obtaining an IR that resembles a realistic RIR. 
Finally, we tested our approach against three baseline methods considering widespread metrics, including reverberation time, clarity, definition, and center time.
Overall, the proposed approach was able to outperform the baseline methods by a large margin across different metrics.

Future work includes the application of the proposed framework to frequency-dependent FDNs, which are able to account for a frequency-specific decay in time, or to multiple-input multiple-output (MIMO) delay networks.

\bmhead{Acknowledgments}
Alessandro Ilic Mezza and Riccardo Giampiccolo wish to thank Xenofon Karakonstantis for the helpful discussion.

\section*{Declarations}
\begin{itemize}
\item Availability of data and materials: 
The datasets 
generated and/or 
analyzed during the current study are available on the MIT Acoustical Reverberation Scene Statistics Survey website, \url{https://mcdermottlab.mit.edu/Reverb/IR_Survey.html}.
\item Funding: This work was partially supported by the European Union under the Italian National Recovery and Resilience Plan (NRRP) of NextGenerationEU, partnership on “Telecommunications of the Future” (PE00000001 --- program “RESTART”), and received funding support as part of the JRC STEAM STM--Politecnico di Milano agreement and from the Engineering and Physical Sciences
Research Council (EPSRC) under the ``SCalable Room Acoustics Modelling (SCReAM)'' Grant EP/V002554/1.
\item Competing interests: The authors declare that they have no competing interests.
\item Authors' contributions: A.\ I.\ Mezza conceptualized the study, designed and implemented the proposed method, run the experiments, and drafted the manuscript. R.\ Giampiccolo contributed to the design of the proposed method, run the experiments, and drafted the manuscript. E.\ De Sena contributed to the method and codebase and drafted part of the manuscript. A. Bernardini revised the manuscript and supervised the work.
\end{itemize}



%%\begin{appendices}


%%\end{appendices}

\bibliography{sn-bibliography}% common bib file

\end{document}